\documentclass[%
twocolumn,
superscriptaddress,
 amsmath,amssymb,
aps,
pre,
10,
]{revtex4-1}

\usepackage{graphicx}
\usepackage{dcolumn}
\usepackage{bm}
\usepackage{hyperref}

\usepackage{color}
\usepackage{natbib}

\begin{document}

\preprint{APS/123-QED}

\title{{First-passage time of run-and-tumble particles with non-instantaneous resetting}}

\author{Gennaro Tucci}
\address{SISSA - International School for Advanced Studies and INFN, via Bonomea 265, I-34136 Trieste, Italy}

\author{Andrea Gambassi}
\address{SISSA - International School for Advanced Studies and INFN, via Bonomea 265, I-34136 Trieste, Italy}

\author{ Satya N. Majumdar}
\address{LPTMS, CNRS, Univ. Paris-Sud, Universit\'{e} Paris-Saclay, 91405 Orsay, France}

\author{Gr\'egory Schehr}
\address{Sorbonne Universit\'e, Laboratoire de Physique Th\'eorique et Hautes Energies, CNRS UMR 7589, 4 Place Jussieu, 75252 Paris Cedex 05, France}

\date{\today}

\begin{abstract}
We study the statistics of the first-passage time of a single run and tumble particle (RTP) in one spatial dimension, with or without resetting, to a fixed target located at $L>0$.
First, we compute the first-passage time distribution of a free RTP, without resetting nor in a confining potential, but averaged over the initial position drawn from an arbitrary
distribution $p(x)$. Recent experiments used a non-instantaneous resetting protocol that 
motivated us to study in particular the case where $p(x)$ corresponds to the stationary non-Boltzmann distribution of an RTP
in the presence of a harmonic trap. This distribution $p(x)$ is characterized by a parameter $\nu>0$, which depends on the microscopic parameters of the RTP dynamics. 
We show that the first-passage time distribution of the free RTP, drawn from this initial distribution, develops interesting singular behaviours, depending on the parameter $\nu$.
We then switch on resetting, mimicked by thermal relaxation of the RTP in the presence of a harmonic trap. Resetting leads to a finite mean first-passage time (MFPT) and
we study this as a function of the resetting rate for different values of the parameters $\nu$ and $b = L/c$ where $c$ is the right edge of the initial distribution $p(x)$. In the diffusive limit of the RTP dynamics, we find a rich phase diagram in the $(b,\nu)$ plane, with an interesting re-entrance phase transition. Away from the diffusive limit, qualitatively similar rich behaviours emerge for the full RTP dynamics.  
\end{abstract}

\maketitle

\section{Introduction}

Search processes appear naturally in a wide range of contexts, such as in animal movements during foraging, biochemical reactions, data search by randomised algorithms and all the way to behavioral psychology -- for a review see Ref.~\cite{benichou2011intermittent}. Finding an optimal search strategy in a given context is fundamental for practical applications. 
Among randomised search strategies, an interesting one involves stopping and restarting the search from scratch at randomly distributed Poissonian times. There has been enormous activities on search processes via stochastic resetting -- see Ref.~\cite{Evans_2020} for a recent review. 
Stochastic resetting has been found to provide an efficient search algorithm in several contexts, such as in optimization algorithms \cite{villen1991restart,luby1993optimal,PhysRevLett.88.178701,tong2008random,avrachenkov2013markov,lorenz2018runtime}, chemical reactions \cite{Reuveni4391,PhysRevE.92.060101}, animal foraging \cite{boyer2014random,majumdar2015random,mercado2018lotka,maso2019anomalous,pal2020search} and catastrophes in population dynamics \cite{levikson1977age,manrubia1999stochastic,visco2010switching}.

{One of the simplest models describing such situations is provided by a particle performing Brownian motion} with stochastic resetting: the particle diffuses with diffusion constant $D$, {and, randomly in time
with constant rate $r$, its position is {\it instantaneously} reset to a {\it fixed location} $x_r$ \cite{Reset,evans2011diffusion}. Resetting of stochastic processes turns out to have two rather generic major consequences: (i) it typically drives the system into a nontrivial stationary state and (ii) in many cases, an optimal resetting
rate emerges which minimises the mean first-passage time (MFPT) of the reset process to a given fixed target. Both aspects have motivated a lot of theoretical work 
during the last few years~\cite{Evans_2020,Evans_2013,PhysRevLett.113.220602,majumdar2015dynamical,Bhat_2016,PhysRevLett.118.030603,PhysRevLett.121.050601,PhysRevE.100.032110,PhysRevLett.120.080601,PhysRevLett.125.050602,PhysRevE.93.062411,PhysRevResearch.2.043138,de2021resetting}.}

\begin{figure}
\centering
\includegraphics[width = 0.8\linewidth]{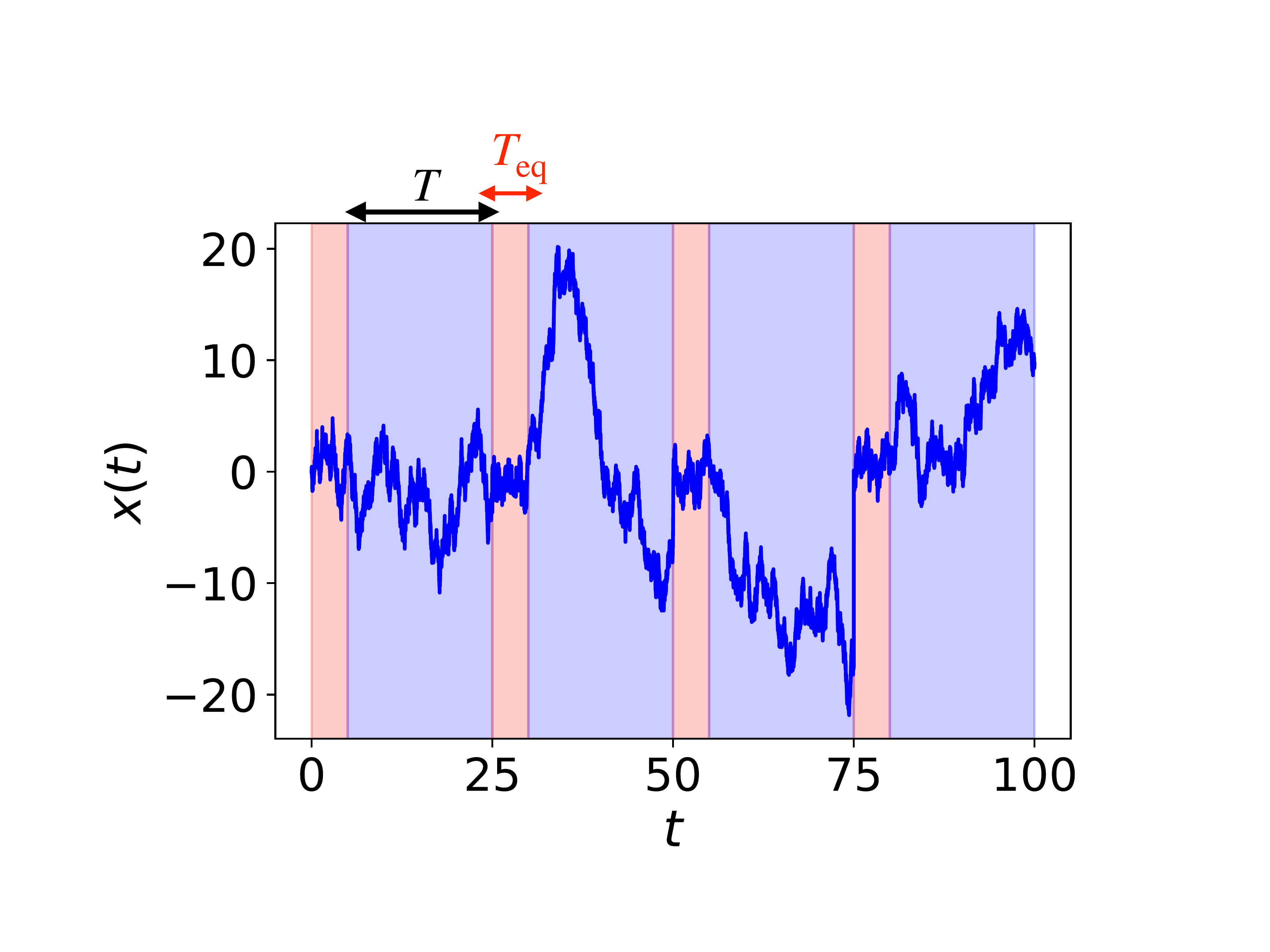}
\caption{A typical trajectory of a Brownian particle evolving according the experimental protocol proposed in Refs. \cite{Ciliberto,Faisant_2021} with periodic resetting. During the \textit{equilibration} phase (red shaded areas) the Brownian particle relaxes in a harmonic trap for a time $T_{\rm eq}$. At the end of this phase, the trap is switched off and the particle diffuses freely (blue shaded area) for a period $T$. For this particular realization of the process we have chosen $T_{\rm eq}=10\,\tau_{\rm rel}$ and $T=40\,\tau_{\rm rel}$, where $\tau_{\rm rel}$ denotes the relaxation time scale of the particle in the trap.} \label{Fig:expBM}
\end{figure}

{It is only recently that resetting protocols were realized experimentally \cite{tal2020experimental,Ciliberto,Besga,Faisant_2021}. These experimental works  revealed two important facts: (i) physical resetting is often non-instantaneous and (ii) it is unrealistic to reset the particle exactly at a fixed position~$x_r$. {Non-instantaneous resetting has been studied in various theoretical models~\cite{reuveni2016optimal,evans2018effects,pal2019time,pal2019invariants,gupta2020work,bodrova2020resetting},
but here we are interested in a particular non-instantaneous resetting protocol
that has been used in recent experiments in optical traps.} In particular, in Refs. \cite{Ciliberto, Besga,Faisant_2021}, the experiments are conducted on colloidal particles diffusing in the presence of a harmonic trap. There were two protocols used for the duration of the free diffusion, either the duration is a fixed period $T$ (periodic resetting) or it is an exponentially distributed random variable (Poisonian resetting). The trap is realized via optical tweezers and it is well approximated by a harmonic potential $V(x) = \kappa x^2/2$. This experimental protocol consists of two distinct phases which alternate in time. First we have an \textit{equilibration} phase of fixed duration $T_{\rm eq}$: the harmonic trap is switched on and the dynamics of the particle reaches a thermal equilibrium at inverse temperature $\beta$. This phase is indicated by the red shaded area in Fig. \ref{Fig:expBM}.
It follows that the position $x$ of the particle is distributed according to the Boltzmann weight $p(x) \propto {\rm e}^{-\beta V(x)}$, with $V(x) = \kappa x^2/2$, i.e., a Gaussian distribution centered at $x=0$ with variance $\sigma^2 \propto 1/\beta$.
Equilibrium is attained if the relaxation time scale of the particle in the trap $\tau_{\rm rel}$ is much smaller than the duration $T_{\rm eq}$ of the equilibration phase.
 At the end of the equilibration phase, the trap is then switched off and the particle diffuses freely during a certain time $T$ (represented by the blue shaded area of Fig. \ref{Fig:expBM}). Then, these two phases are repeated cyclically.
 
 For $T_{\rm eq}\gg\tau_{\rm rel}$, we see that this setup mimics a non-instantaneous resetting protocol where the particle is reset to a random position $x_r$. One important feature of this protocol is that $x_r$ is itself drawn randomly from a certain probability distribution function (PDF) $p(x)$, which in the case of a Brownian particle is the Boltzmann distribution $p(x) \propto {\rm e}^{-\beta V(x)}$.} 
 
{One of the main focuses of Refs.
\cite{Ciliberto, Besga,Faisant_2021} was on the first-passage time (FPT) distribution to a target located at $x=L$.
Note that no measurement was performed during the equilibration phase -- this is meant to reproduce instantaneous resetting. Moreover, the experimental protocol presented above can be easily adapted to the case of no resetting, i.e., by taking the limit $T\rightarrow\infty$  \cite{Ciliberto, Besga,Faisant_2021}. In this case, a natural question is thus what happens to the FPT distribution to a fixed target located at $L$ after {\it averaging} over the initial position, distributed
with a certain PDF $p(x)$, the stationary distribution corresponding to the external potential $V(x)$. In the case of Brownian motion, for which $p(x)$ is simply the Boltzmann weight $p(x) \propto {\rm e}^{-\beta V(x)}$, i.e., in this case, a Gaussian of zero mean and variance $\sigma^2 \propto 1/\beta$, it was shown that the averaged FPT exhibits a very rich behavior, including a dynamical phase transition between a two-peaked and a one-peaked shape as the ratio $L/\sigma$ is varied \cite{Besga}. This transition was not only predicted theoretically but also observed in experiments \cite{Besga}. Given the relevance of FPT for a variety of applications in physics literature \cite{redner2001guide,bray2013persistence}, it is then natural to extend these studies to other stochastic processes, beyond the simple Brownian motion.}

In this {paper, we study the one-dimensional persistent random walk, also known as the run-and-tumble particle (RTP). The dynamics of this model consists of two alternating phases: running and tumbling. During the running phase, the particle moves ballistically with a fixed velocity $v$, during an {\it exponentially} distributed random time with mean $1/\gamma$. At the end of the running phase, the particle tumbles instantaneously and chooses a new direction for the next running phase.} This simple model has been used to describe the motion of some species of bacteria, e.g.,  \textit{Escherichia coli} \cite{Coli,PhysRevLett.100.218103}. In these cases, the bacteria self-propel by consuming energy directly from the environment \cite{PhysRevLett.100.218103,nash2010run,elgeti2015run,solon2015active,cates2013active}. The existence of a finite run-time induces a memory in the RTP dynamics, rendering it non-Markovian, as opposed to the Markovian Brownian motion. The dynamics of a free RTP has been studied extensively and many exact results are known \cite{orsingher1990probability,hanggi1995colored,weiss2002some,Malakar_2018,de2021survival,angelani2015run}. Very recently, for the RTP,  
{the effect of {\it instantaneous} resetting to a fixed location $x_r$ has been studied in Refs. \cite{Evans_2018,masoliver2019telegraphic,Bressloff_2020,santra2020run}. It was found that, as in the Brownian case, resetting drives an RTP into a non trivial stationary state and can optimize the MFPT to a fixed target. 

As discussed above, it is very hard to achieve experimentally the instantaneous resetting usually assumed in theoretical settings \cite{Ciliberto, Besga,Faisant_2021}. Typical experimental protocols, used for Brownian particles, involve switching on and off the trap and letting the particles equilibrate in between. One of the main effects of this protocol corresponds to choosing 
the resetting position $x_r$ randomly from the stationary distribution inside the trap, which for the Brownian case, happens to be the equilibrium Boltzmann distribution. 
It is then natural to ask: what is the corresponding effect of this experimental protocol in the case of RTP, where the stationary distribution is known to be non-Boltzmann? This is the main 
question that we address in this paper. Moreover, in the case of an RTP in an harmonic trap, the non-Boltzmann stationary distribution has a finite support/width. We demonstrate that the combined effect of the {\it finite width} of the stationary distribution and the resetting leads to a rather rich and interesting physics. We expect that the results presented in this paper will be useful for possible future experimental investigations of RTP with resetting.

The rest of the paper is organised as follows.
In Section \ref{sec:model}, we recap some results on the FPT {of a free RTP as well as some properties of the stationary state of an RTP in the presence of an external confining potential $V(x)$.}
In Section \ref{sec:RT}, we consider the case {\it without resetting} and compute the FPT distribution of a free RTP averaged over the initial position drawn from the stationary distribution of the RTP in the presence of a harmonic trap. In Section \ref{sec:RT_reset}, we study how stochastic resetting affects the MFPT for the particles. In Section \ref{sec:RT_periodic},
we extend our results to the periodic resetting protocol. Finally, we conclude in Section~\ref{sec:Concl}. Some details of the computations are relegated to the appendices.

\begin{figure}
\centering
\includegraphics[width = 0.8 \linewidth]{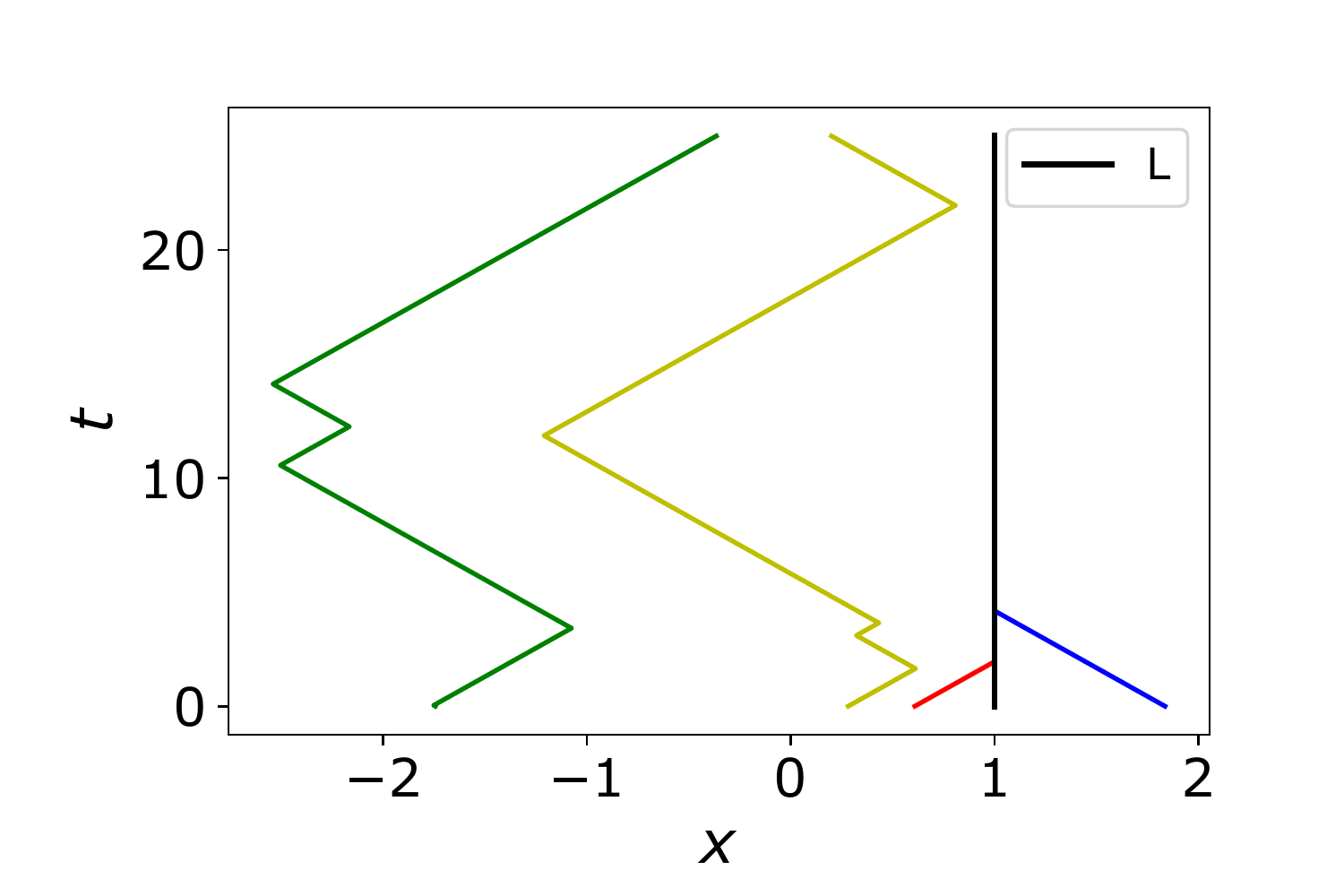}\\
\caption{Typical trajectories of RTPs, in the presence of an absorbing boundary located at $L  = 1$ (shown by a solid black vertical line). The two rightmost trajectories got already absorbed at $L$ within the time shown, while the two leftmost trajectories are yet to be absorbed. We chose the parameters to be $\gamma=0.15$, $v=0.2$ and $\nu = 1.5$.}\label{Fig:p1}
\end{figure}

\section{{One-dimensional RTP with and without an external potential: a reminder}}\label{sec:model}
The position $x(t)$ of a one-dimensional RTP starts from $x(t=0)=x_0$ and then evolves according to the following Langevin equation \cite{kac1974stochastic}
\begin{equation}\label{eq:RTLangevin}
\dot{x}(t)=v \sigma_0(t) \;,
\end{equation}
where $\sigma_0(t)$ is a telegraphic noise which switches between $+1$ and $-1$ with rate $\gamma$, while $v>0$
is the modulus of the velocity of the particle (which is fixed here). We now {assume that there is an absorbing target at position $L$: typical trajectories of the process are represented in Fig. \ref{Fig:p1}). We denote by $S_0(t|d_L)$ (respectively $F_0(t|d_L) = -\partial_t S_0(t|d_L)$) the survival probability at time $t$ (respectively the FPT distribution) given that the particle started at a distance $d_L(x_0) \equiv |L-x_0|$ from the target. These observables have been widely studied in the literature \cite{orsingher1990probability,hanggi1995colored,weiss2002some,Malakar_2018,de2021survival,angelani2015run}. The Laplace transform of the survival probability $S_0(t|d_L(x_0))$ reads
\begin{eqnarray} \label{EqS0}
\tilde S_0(s|d_L(x_0)) &=& \int_0^\infty  {\rm d}t \, {\rm e}^{-st} S_0(t|d_L(x_0)) \\
&=& \frac{1}{s}\left[1+\frac{v\lambda(s)-s-2\gamma}{2\gamma}\,{\rm e}^{-\lambda(s)d_L(x_0)}\right] \;, \nonumber
\end{eqnarray}
where 
\begin{eqnarray} \label{eq:lambda}
\lambda(s) \equiv \frac{\sqrt{s^2+2 \gamma s}}{v} \;.
\end{eqnarray}
It is actually more convenient to consider the FPT distribution, for which the Laplace inversion can be explicitly carried out and it reads, in dimensionless units, 
\begin{eqnarray}
F_0(t|d_L(x_0)) &=& \gamma f_0\left(\tau \equiv \gamma t| y\equiv {\gamma} \frac{d_L(x_0)}{v}\right) \;, \nonumber \\
{\rm with}\quad f_0(\tau|y)&=&\frac{{\rm e}^{-\tau}}{2}\left[\delta(\tau-y)+\theta(\tau-y)g_0(\tau|y)\right] \label{eq:f0}
\end{eqnarray}
where $\theta(z)$ denotes the Heaviside step function. The function $g_0(\tau|y)$ reads 
\begin{equation}\label{eq:g0}
\begin{aligned}
g_0(\tau|y)&\equiv \frac{y}{y+\tau}I_0\left(\sqrt{\tau^2-y^2}\right)\\
&+\left[\frac{\sqrt{\tau-y}}{(\tau+y)^{\frac{3}{2}}}+\frac{y}{\sqrt{\tau^2-y^2}}\right]I_1\left(\sqrt{\tau^2-y^2}\right),
\end{aligned}
\end{equation}
where $I_n(z)$ denotes the modified Bessel function of the first kind of index $n$ \cite{abramowitz}.
The first term in Eq. \eqref{eq:f0}, proportional to $\delta(\tau-y)$, accounts for trajectories that reach the target located at $L$ without tumbling, which occurs with probability ${\rm e}^{-\tau}$ in rescaled units. The factor $1/2$ comes from the initial condition, where the initial velocity is $\pm v$ with equal probability. 
In contrast, the second term comes from trajectories with at least one tumbling event. The function $\theta(\tau-y)$ in the second contribution to $f_0(\tau|y)$ in Eq. (\ref{eq:f0}) expresses the fact that particles need a minimal (dimensionless) time $\tau_{0}=y$ to reach the target at $L$. For large time $\tau$, using the asymptotic behavior of the Bessel function $I_n(z)\simeq (2\pi z)^{-1/2} \, {\rm e}^z [1+O(z^{-1})]$ for large $z$, one finds that $g_0(\tau|y)$ in Eq. (\ref{eq:g0}) behaves, when $\tau \to \infty$, as
\begin{eqnarray}\label{g0_asympt}
g_0(\tau|y) \simeq \frac{2 \, {\rm e}^\tau}{\sqrt{2 \pi \tau^3}}(y+1/2) \;.
\end{eqnarray}
By substituting this asymptotic behavior (\ref{g0_asympt}) in Eq. (\ref{eq:f0}), one obtains the 
(scaled) FPT distribution $f_0(\tau|y)$ in Eqs.~(\ref{eq:f0}) and (\ref{eq:g0}) for large $\tau$ as
\begin{eqnarray}\label{large_tau}
f_0(\tau|y)\simeq \frac{(y+1/2)}{\sqrt{2\pi\tau^3}} \;.
\end{eqnarray}
Its long-time algebraic decay $\propto \tau^{-3/2}$ coincides with that of a free Brownian motion, albeit with a different amplitude. In particular, the amplitude does not vanish as $y \to 0$ \cite{le2019noncrossing}. The 
expression for the Brownian motion is recovered in the scaling limit
\begin{eqnarray}\label{diff_lim} 
 v \to \infty \;,\; \gamma \to \infty \quad {\rm with}\quad \frac{v^2}{2 \gamma} = D \quad {\rm fixed} \;. 
 \end{eqnarray}
 In this limit, one finds indeed that $F_0(t|d_L(x_0)) \to F_{\rm BM}(t|d_L(x_0))$ where 
 \begin{eqnarray} \label{BM_result}
 F_{\rm BM}(t|d_L(x_0))=\frac{d_L(x_0)}{\sqrt{
4\pi D t^3}}\,\exp\left(-d_L^2(x_0)/(4Dt)\right).
\end{eqnarray}
Thus one recovers the well known result for the Brownian motion~\cite{redner2001guide,bray2013persistence}. 

\begin{figure}[t]
\centering
\includegraphics[width = 0.9 \linewidth]{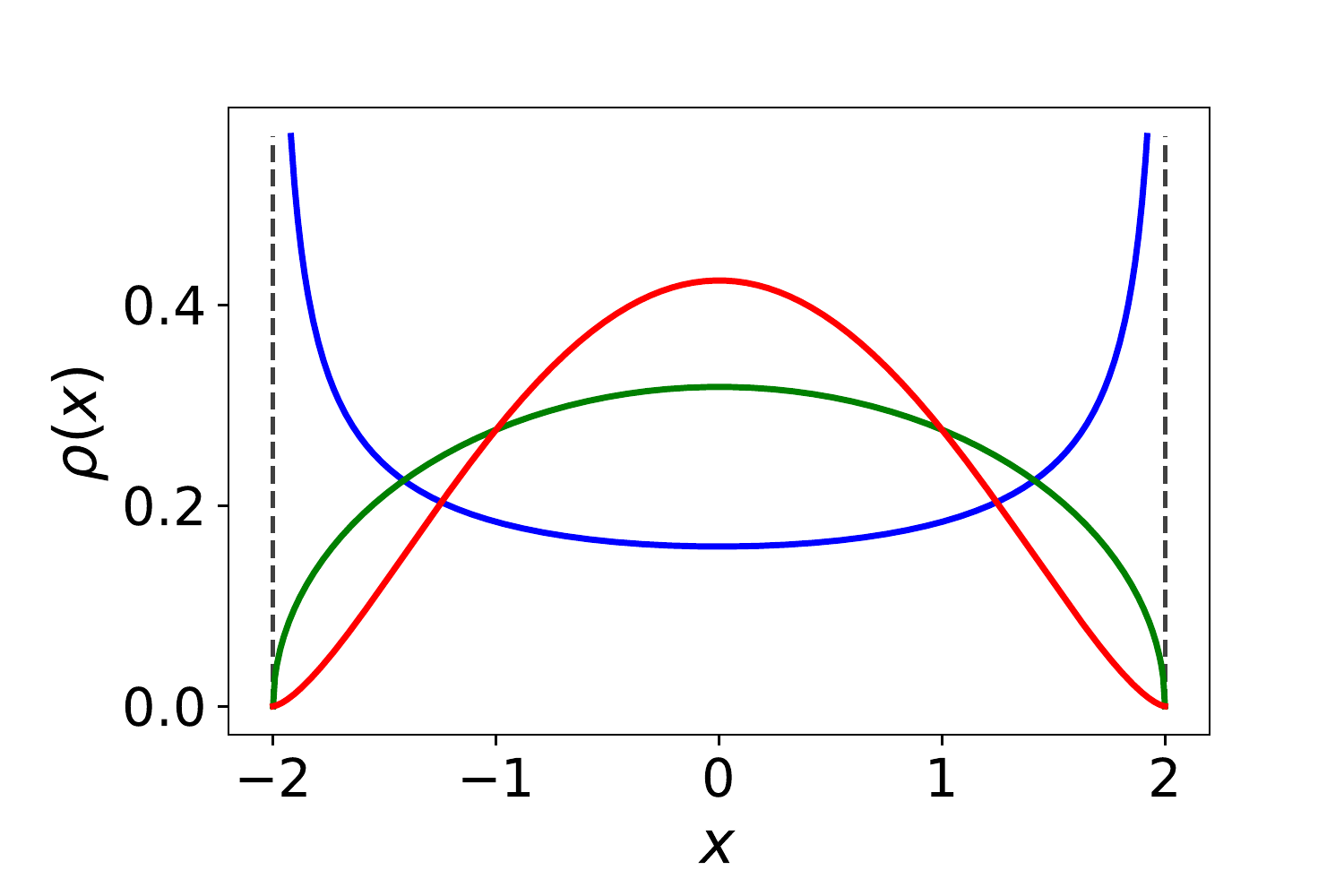}
\caption{Plot of the probability distribution $p(x)=\rho(x/c)/c$ in Eq. \eqref{eq:rho} with $c=2$ and different choices of $\nu$. \textit{Passive} regime: the red and green lines correspond respectively to $\nu=2.5$, and $\nu=1.5$. \textit{Active} regime: the blue line corresponds to $\nu=0.5$.}\label{Fig:p}
\end{figure}

The properties discussed so far are relevant to describe the dynamics of the RTP during the phases where the external potential is switched off, such that the RTP moves freely as in Eq.  (\ref{eq:RTLangevin}). What happens when the external potential $V(x)$ is turned on? During this phase, the dynamics of the RTP is described by the overdamped Langevin equation
\begin{eqnarray} \label{Langevin:pot}
\dot x(t) = - V'(x) + v \sigma_0(t) \;.
\end{eqnarray}
Interestingly, it was shown that, rather generically (namely if $V(x)$ is sufficiently confining), the RTP will converge
to a stationary state. In this stationary state, the PDF of the position of the RTP can be calculated explicitly \cite{klyatskin1977dynamic,kitahara1980phase,hanggi1995colored,solon2015pressure,Confpot,demaerel2018active}.  
In particular, in the case of the harmonic potential $V(x)=\kappa x^2/2$, the stationary state is characterised by three parameters
\begin{equation} \label{para}
\gamma \; ({\rm flip \; rate})\,, \; v \; ({\rm intrinsic \; speed}) \; {\rm and} \; \kappa \; (\rm{trap \, stiffness}) \;.
\end{equation}
The stationary PDF $p(x)$ has a finite support $[-c,+c]$, with $c=v/\kappa$~\cite{Confpot,demaerel2018active}
\begin{eqnarray}\label{eq:rho}
p(x) &=& \frac{1}{c}\,  \rho\left(\frac{x}{c} \right) \;, \nonumber \\
{\rm where} \quad  \; 
    \rho(z)&=&\theta(1-z^2)\,N(\nu)\,(1-z^2)^{\nu-1},
\end{eqnarray}
with $\nu=\gamma/\kappa$ and the normalization constant $N(\nu)$ is given by  
\begin{eqnarray} \label{Nu}
N(\nu) =  \frac{\Gamma\left(\nu+1/2\right)}{\sqrt{\pi}\,\Gamma\left(\nu\right)} \;.
\end{eqnarray}
Physically, the location of the edges $\pm c$ of the support of the distribution $p(x)$ correspond to the points where the velocity of the particle vanishes, i.e., $V'(\pm c)=\pm v$. Near the edges, 
as $x \to \pm c$, the PDF $p(x)$ behaves as $p(x) \propto (c-|x|)^{\nu-1}$. This indicates that $p(x)$ exhibits a qualitative change as $\nu$ crosses the value $\nu =1$ \cite{Confpot}. For $\nu>1$, $p(x)$ is bell-shaped and vanishes at the edges $\pm c$: this case is called {\it passive} \cite{Confpot} since this bell-shaped curve is qualitatively similar to the Gaussian distribution corresponding to a Brownian passive particle. Note also that for $\nu>2$, the slope of $p(x)$ at $x=\pm c$ is finite, while it diverges for $1<\nu<2$, corresponding, respectively, to the red and the green curves in Fig. \ref{Fig:p}. On the other hand, for $0<\nu<1$, the particle tends to accumulate at the edges of the distribution: this results in a U-shaped $p(x)$ that diverges at $\pm c$, see the blue curve of Fig. \ref{Fig:p} -- this is the {\it active case}~\cite{Confpot}.


\section{The FPT distribution for a free one-dimensional RTP averaged over initial condition}\label{sec:RT}

\begin{figure}[t]
\includegraphics[width = \linewidth]{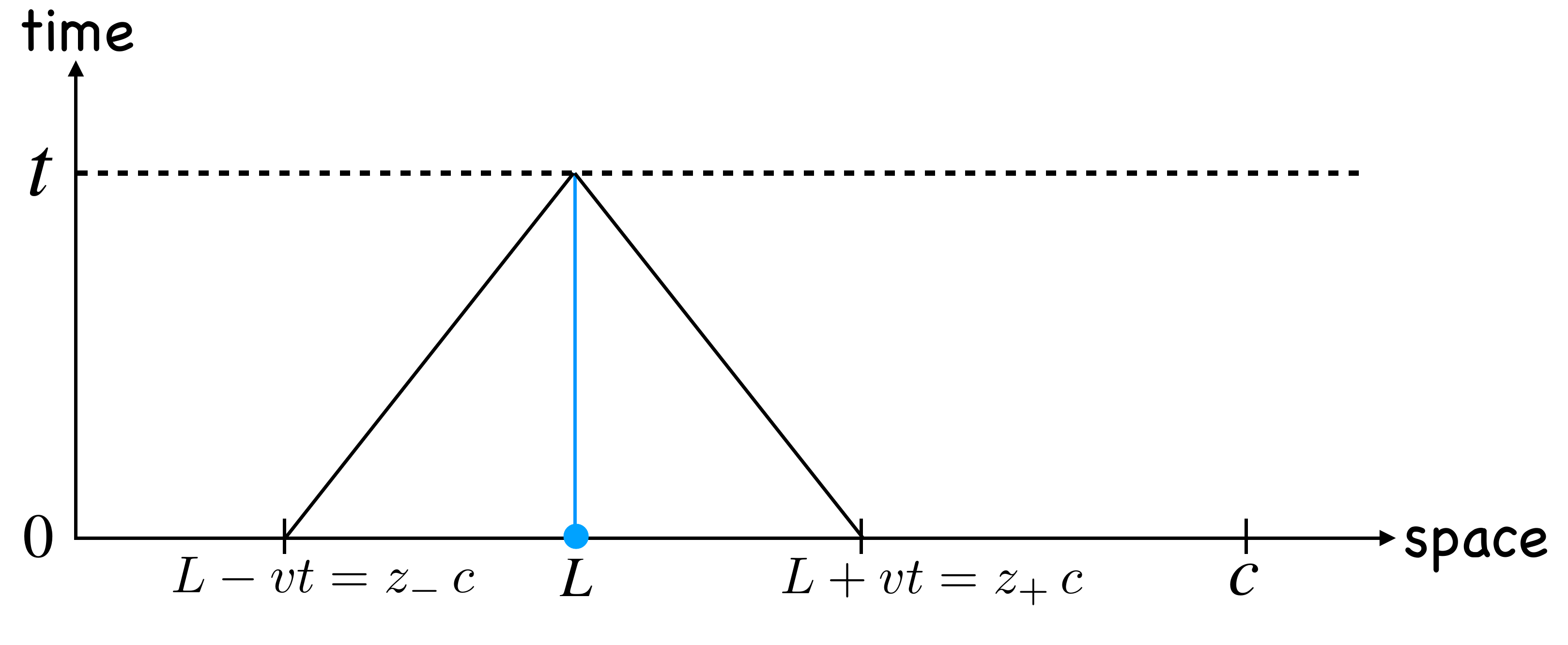}
\caption{Schematic space-time trajectories of the RTP that starts at $t=0$ and reaches $L$ for the first time at time $t$ without tumbling. The trajectory can either arrive from the left or from the right of $L$, as represented by the two straight lines. In order to arrive from the right (respectively from the left), the trajectory must have started at $L+vt = z_+\,c$ (respectively from $L-vt = z_-\,c$).}\label{Fig_space_time}
\end{figure}

In this section, motivated by the experimental resetting protocol discussed in the introduction, we first consider the distribution $F(t)$ of the FPT of a free run and tumble particle to a target located at $L$, averaged over
the initial position. This reads
\begin{eqnarray} \label{av_FPT}
F(t) = \int_{-\infty}^{\infty} \mathrm{d} x_0 \, p(x_0) \, F_0(t|d_L(x_0)) \;,
\end{eqnarray}
where $F_0(t|d_L(x_0))$ is the FPT distribution for a free RTP starting at a certain distance $d_L(x_0) = |L-x_0|$ from the target given in Eq.~(\ref{eq:f0}). Here, $p(x_0)$ is the stationary PDF of the position of an RTP in the presence of an external potential $V(x) = \kappa\,x^2/2$, given in Eq. (\ref{eq:rho}).

It is convenient to express $F(t)$ in terms of the dimensionless variables
\begin{equation}\label{eq:alfa}
a = \frac{\gamma L}{v}
\;\;\;\text{and}\;\;\;b = \frac{L}{c}\;, 
\end{equation}
where $a$ represents the ratio between the position of the target $L$ and the typical distance $l_\gamma =  v/\gamma$ travelled by the particle between two consecutive tumblings, while $b$ expresses how far the target is compared to the size $\sim c$ of the support of $p(x)$. To simplify the discussion, we choose $L \geq 0$ (by symmetry the case $L\leq 0$ can be treated in the same way). Substituting Eqs. (\ref{eq:f0}) and (\ref{eq:g0}) in Eq. (\ref{av_FPT}), we get
\begin{eqnarray}
F(t) &=& \gamma f(\tau = \gamma t) \;, \; \nonumber\\
f(\tau) &=&\frac{{\rm e}^{-\tau}}{2}\left\{\frac{b}{a}\left[\theta\left(1-|z_+|\right)\rho\left(z_+\right)+\theta\left(1-|z_-|\right)\rho\left(z_-\right)\right]\right.\nonumber \\
&&\left.+\int_{\max(-1,z_-)}^{\min(1,z_+)}\mathrm{d}z\,\rho(z)g_0\left(\tau |\,a \,l(z)\right)\right\}\;, \label{eq:f} 
\end{eqnarray}
where we have introduced the dimensionless variables
\begin{eqnarray} \label{def_l}
z_{\pm} &=& b\left(1 \pm\frac{\tau}{a} \right) = \frac{L \pm v\,t}{c} \;, \\
{\rm and}\quad l(z)&=&  \Big|\frac{z}{b}-1\Big| \;, 
\end{eqnarray}
while the expression of $g_0(\tau|y)$ is given in Eq.~\eqref{eq:g0}.
The first two terms in the expression of $f(\tau)$ in Eq.~\eqref{eq:f} correspond to particles that have reached the target for the first time at the (scaled) time $\tau$ without experiencing a tumble, which occurs with a probability $\propto {\rm e}^{-\tau}$. The first term $\propto \theta\left(1-|z_+|\right)\rho\left(z_+\right)$ corresponds to particles having a velocity $-v$ (which thus started from the initial position $L+v\,t$). The second one $\propto \theta\left(1-|z_-|\right)\rho\left(z_-\right)$ corresponds to particles having a velocity $+v$ (which thus started from the initial position $L-v\,t$) -- see Fig.~\ref{Fig_space_time}. The third and last term in 
Eq. (\ref{eq:f}) corresponds to particles that reached the target for the first time at $\tau$, having experienced at least one tumble. It turns out that 
this last term controls the long-time asymptotic behavior of $f(\tau)$. To derive this asymptotic behavior of $f(\tau)$, we use in the third term in Eq. (\ref{eq:f})
the expression for $g_0(\tau|y)$ from Eq. (\ref{g0_asympt}) for large $\tau$. This gives
\begin{eqnarray}\label{eq:frt0}
f(\tau)\simeq \frac{1}{\sqrt{2\pi\tau^3}} \left[\frac{1}{2}+a\int_{-1}^{+1}\mathrm{d}z\,\rho(z)\,l(z) \right] \;.
\end{eqnarray}
\begin{figure}[t]
\includegraphics[width = \linewidth]{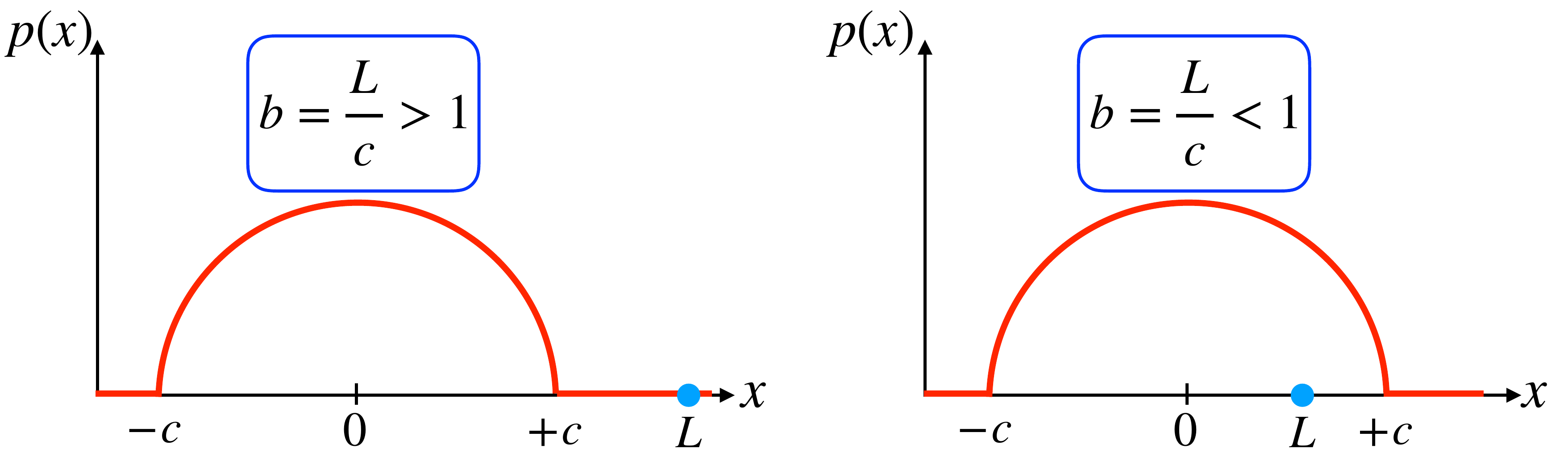}
\caption{Schematic representation of the position of the target (at $L>0$) relative to the right edge (at $+c$) of the stationary RTP distribution $p(x)$ with $\nu = 3/2$. On the left panel, the target is outside the support, corresponding to $b=L/c>1$, while on the right panel, the target is inside the support, corresponding to $b=L/c<1$.} \label{Fig_target}
\end{figure}
For $\rho(z)$ in Eq. \eqref{eq:rho}, one has
\begin{equation}\label{eq:ampl}
\begin{aligned}
&\int_{-1}^{+1}\mathrm{d}z\,\rho(z)\,l(z)=\\
&\begin{cases}
1 & \mbox{for}\,\, b>1,\\
N(\nu)\left[2b\, {}_2F_1\left(\frac{1}{2},1-\nu,\frac{3}{2},b^2\right)+\frac{(1-b^2)^\nu}{\nu b}\right]& \mbox{for}\,\, b<1,
\end{cases}
\end{aligned}
\end{equation}
where ${}_2F_1$ denotes the hypergeometric function and $N(\nu)$ is given in Eq. (\ref{Nu}). The FPT in Eq. \eqref{eq:frt0} exhibits a standard $\tau^{-3/2}$ decay -- as in the Brownian case -- albeit with a different amplitude that depends on the distribution $p(x)$.  
In the opposite limit of shorter times, the average FPT distribution $F(t)$ develops singularities which are due to the two first terms in Eq. \eqref{eq:f}. These singularities arise because of the theta-functions and they thus occur for $|z_\pm|=1$. Their nature differs depending on whether the target $L$ is inside the support $[-c,+c]$ of $p(x)$ (corresponding to $b < 1$) or outside it (corresponding to $b > 1$) -- see Fig.~\ref{Fig_target} for a schematic illustration. We thus discuss these two cases separately.

\vspace*{0.5cm}
\noindent{\it (i) The case $0<b<1$,} the target is inside the support of $p(x)$, i.e., $0\leq L<c$. Consequently, there are trajectories that hit the target exactly at time $t=0$ (namely the trajectories that start from $x=L$). Hence the left edge of the support of $f(\tau)$ is $\tau_{0}=0$. The theta functions in Eq. \eqref{eq:f} indicate that $f(\tau)$ is singular at $|z_\pm(\tau_\mp)|= 1$, i.e., at
\begin{eqnarray}\label{tau_pm}
\tau_\pm=a\left(\frac{1}{b}\pm 1\right) = \gamma \frac{c \pm L}{v} \;.
\end{eqnarray}
\begin{figure}[t]
\includegraphics[width=\linewidth]{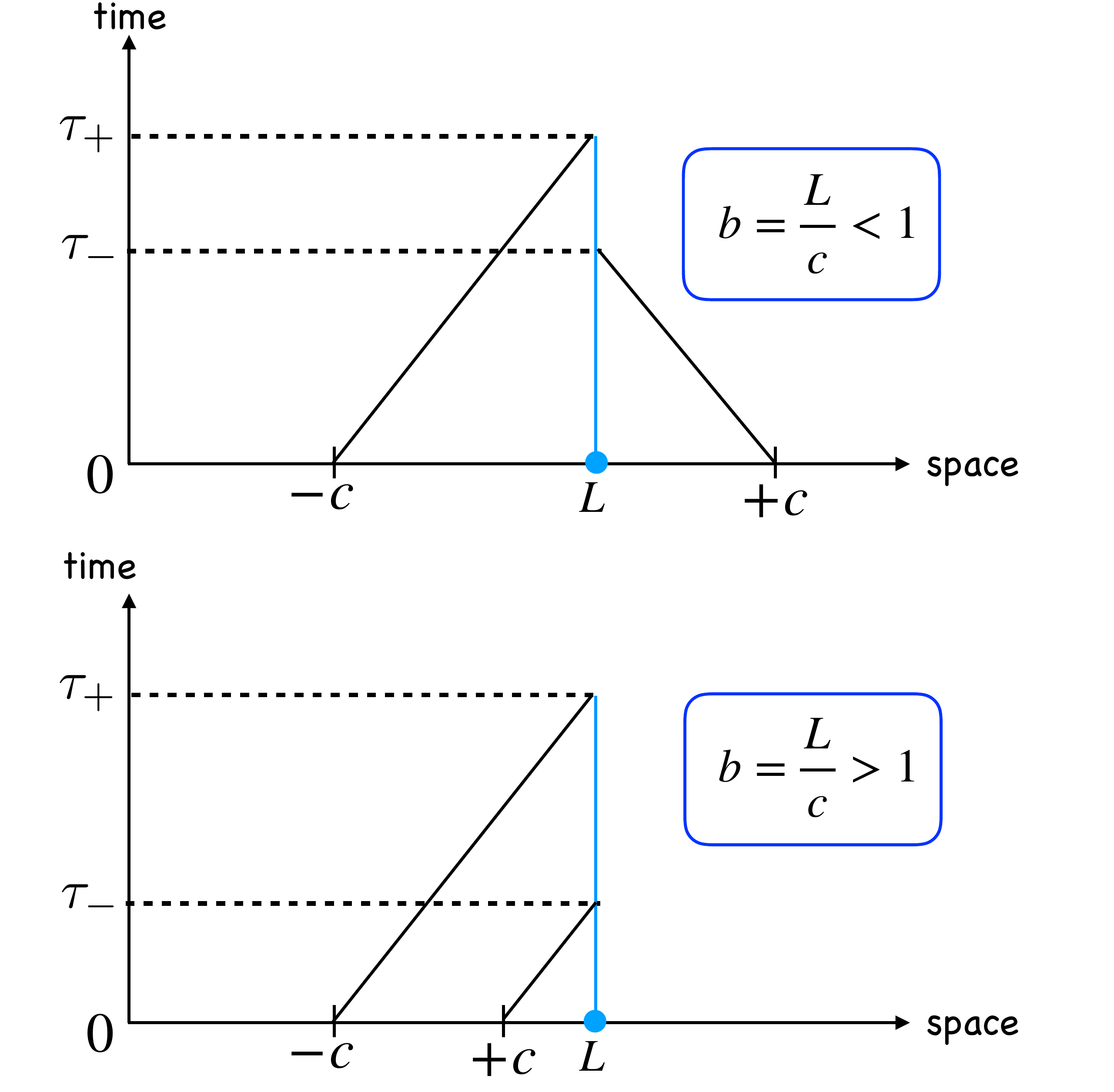}
\caption{Schematic space-time pictures representing the RTP trajectories corresponding to the first two terms on the right hand side of Eq.~(\ref{eq:f}). In the top panel ($b<1$), the left straight line corresponds to the trajectories of the RTP starting from the left edge of the support at $-c$ and arriving for the first time at $L$ at time $\tau_+$, without undergoing any tumbling in-between. The right straight line corresponds to a RTP that starts at the right edge $+c$ and arrives, without tumbling, at $L$ at time $\tau_-$ for the first time. We use the notation $\tau_-$ and $\tau_+$ to indicate that $\tau_- < \tau_+$ (for $L>0$). In the bottom panel ($b>1$), the two lines again correspond to the trajectories that start from the edges $\pm c$ of the support and reach $L$ without tumbling for the first time at time $\tau_{\mp}$.}\label{Fig_space_two_time}
\end{figure}
Clearly, $\tau_+$ (respectively $\tau_-$) is the time needed for a particle moving ballistically with velocity $+v$ (respectively $-v$) to reach the target located at $L<c$ -- see the top panel of Fig. \ref{Fig_space_two_time}. The behavior of $f(\tau)$ close to these singular points depends on the one of $p(x)$ close to the edges at $x = \pm c$. One finds indeed, that for small $\epsilon > 0$
\begin{equation}\label{eq:scaling}
\begin{aligned}
f(\tau_{\pm} - \epsilon)&=\frac{{\rm e}^{-\tau_\pm}}{2}\frac{b}{a}\,\rho\left(1-\epsilon \frac{b}{a}\right)+O(\epsilon^0)\\ 
&= \frac{{\rm e}^{-\tau_\pm}}{4}\left(\frac{2b}{a}\right)^\nu N(\nu)\, \epsilon^{\nu - 1}+O\left(\epsilon^0\right)\,,
\end{aligned}
\end{equation}
where we recall that $N(\nu)$ is given in Eq. (\ref{Nu}). Hence, in the  \textit{active} regime $0<\nu<1$, $f(\tau)$ diverges on the left of $\tau_{\pm}$: this behavior is displayed by the blue curve of Fig. \ref{Fig3}a. 
This divergence comes from particles that are initially located near the edges, where $p(x)$ exhibits a divergence in this case (see Fig. \ref{Fig:p}), and reach the target from $\mp c$ at the time $\tau_\pm$ without tumbling. 
In the \textit{passive} regime $\nu>1$, $f(\tau)$ is continuous at $\tau_\pm$ but it still exhibits a singular behavior close to these points $\tau_{\pm}$. For instance, for $1<\nu<2$ its first derivative diverges on the left of $\tau_{\pm}$, while for $\nu\ge 2$ it is also continuous: this is displayed, respectively, in the green and red curve of Fig. \ref{Fig3}a . {Note that the average FPT is always finite at $\tau_\pm +\epsilon$, i.e.,  
\begin{equation}\label{eq:ampl2}
\begin{aligned}
f(\tau_++\epsilon)=&\frac{{\rm e}^{-\tau_+}}{2}\int_{-1}^{+1}\mathrm{d}z\,\rho(z)g_0(\tau_+|l(z))+O(\epsilon),\\
f(\tau_-+\epsilon)=&\frac{{\rm e}^{-\tau_-}}{2}\left[\int_{2b-1}^{+1}\mathrm{d}z\,\rho(z)g_0(\tau_-|l(z))\right.\\
&\left.\;\;\;\;\;\;+\rho(2b-1)+O(\epsilon)\right],\\
\end{aligned}
\end{equation}
in agreement with the fact that the particles reaching the target at $\tau_\pm +\epsilon$ have experienced at least one tumble.} {At large times, the FPT distribution scales as $\tau^{-3/2}$ and the corresponding  prefactor in Eq. \eqref{eq:frt0} depends on the behavior of $p(x)$ at its boundaries and the location of the target, namely on $\nu, c$ and $L$. In Fig. \ref{Fig3}a, on the scale used here, the dependence of the function $f(\tau)$ for $\tau > \tau_+$ on these parameters is hardly visible. However, we have checked carefully that the differences show up by zooming in on this region. By integrating the FPT distribution in Eq. (\ref{eq:f}), we also computed the survival probability up to time $t$, namely $S(t) = \int_t^{\infty} {\rm d}t' F(t')$. This is plotted in Fig. \ref{Fig3}b where it is compared to numerical simulations, finding excellent agreements.}
\begin{figure*}[ht]
\centering
\includegraphics[width = 0.4 \linewidth]{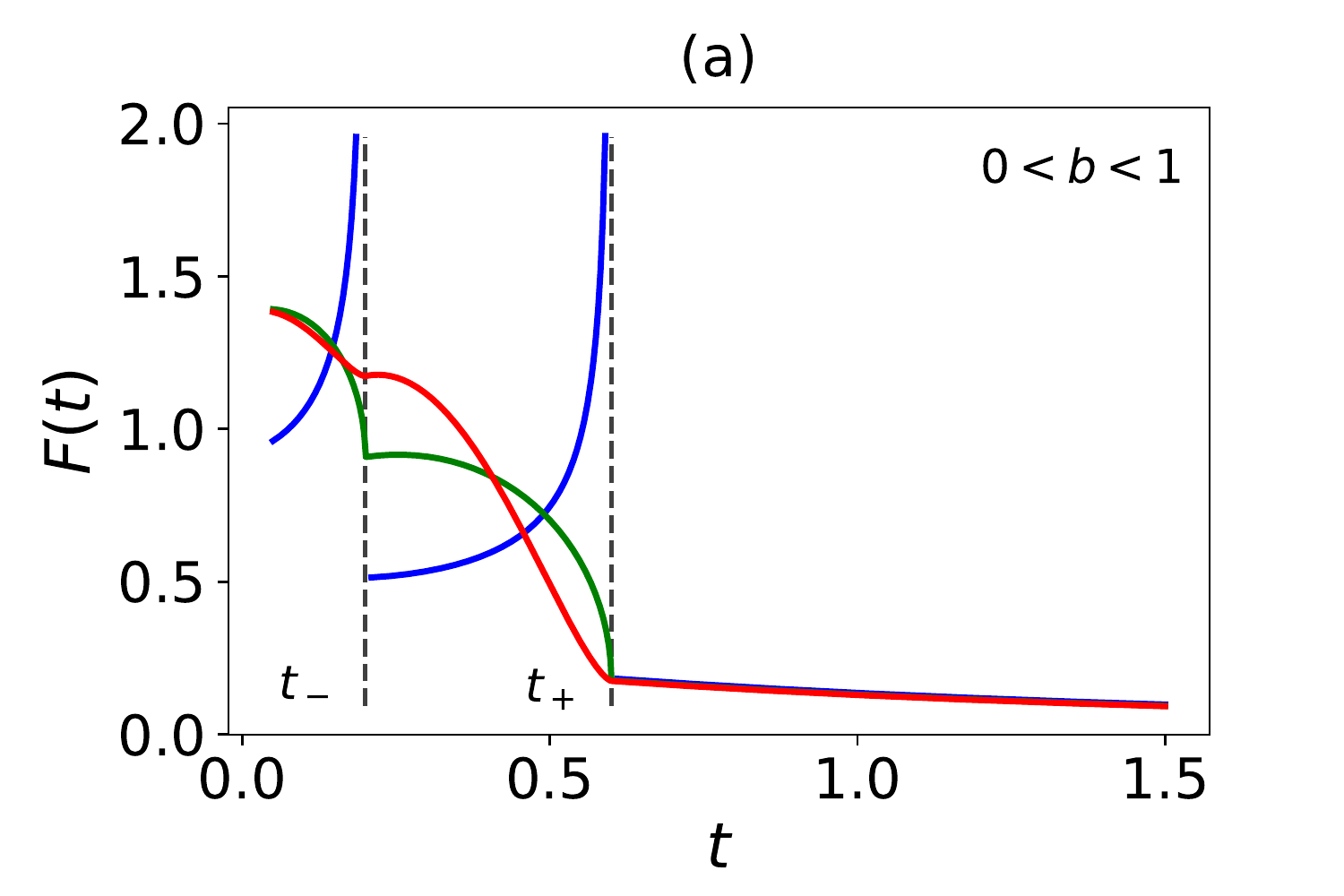}\quad
\includegraphics[width = 0.4 \linewidth]{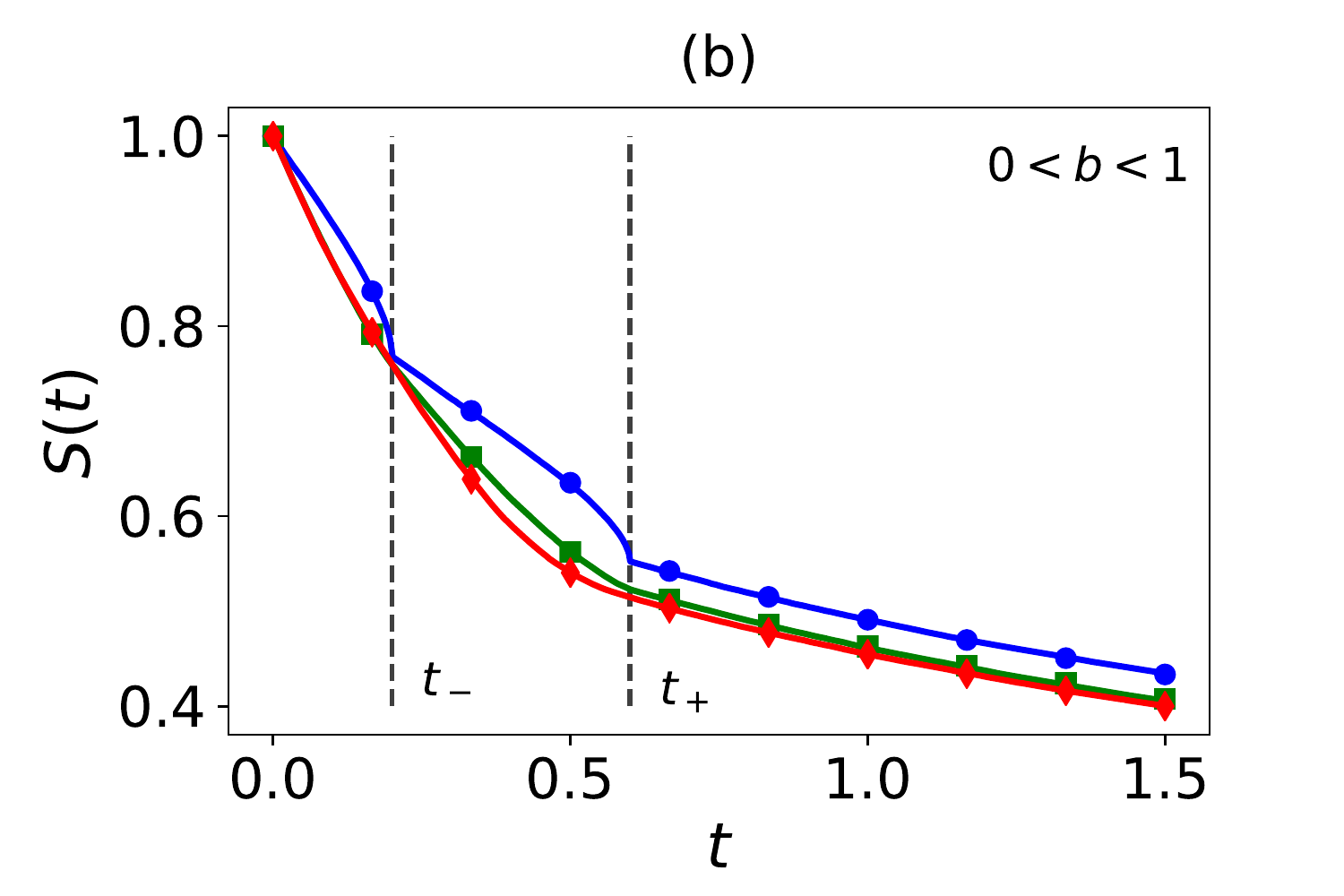}\\
\includegraphics[width = 0.4 \linewidth]{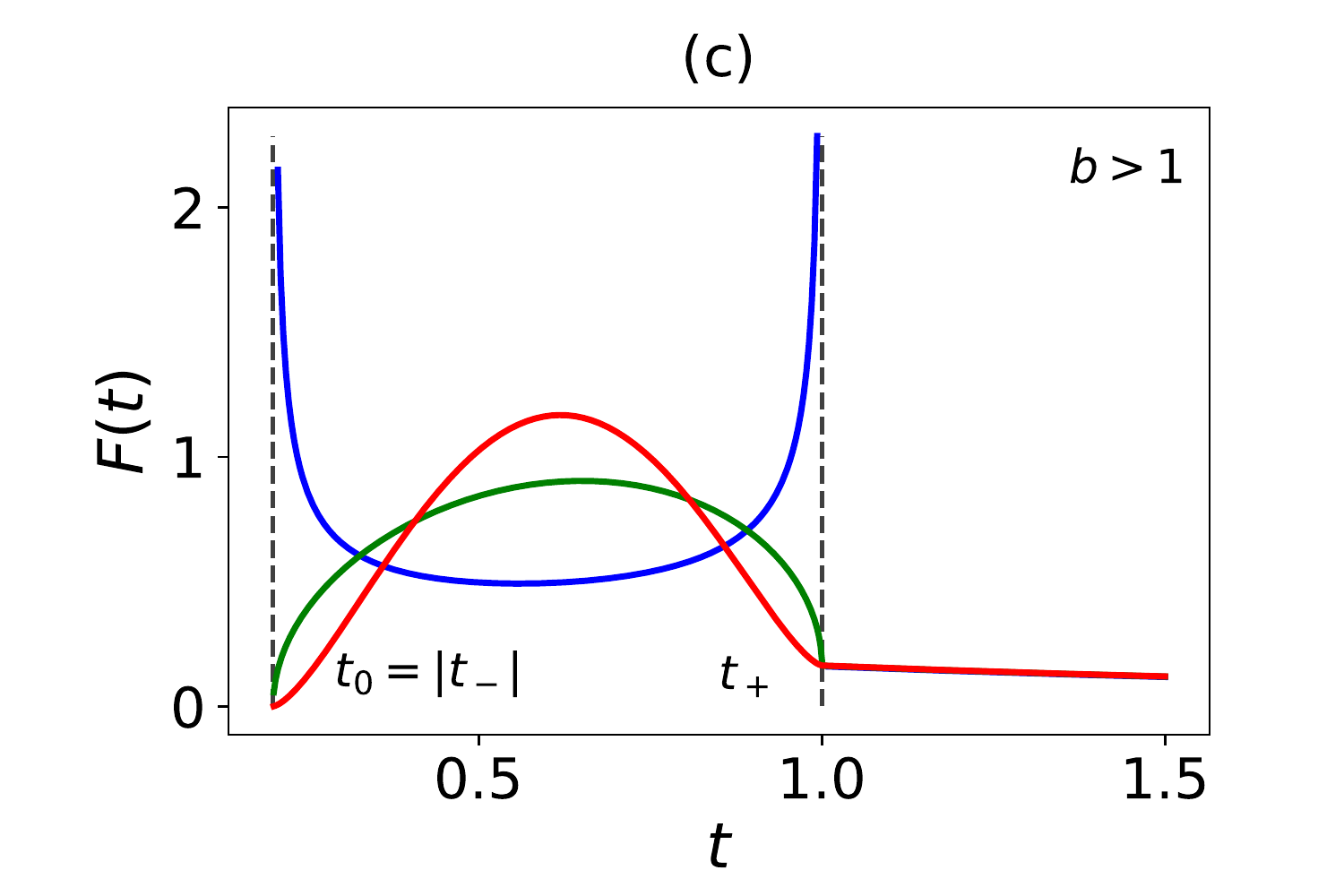}\quad
\includegraphics[width = 0.4 \linewidth]{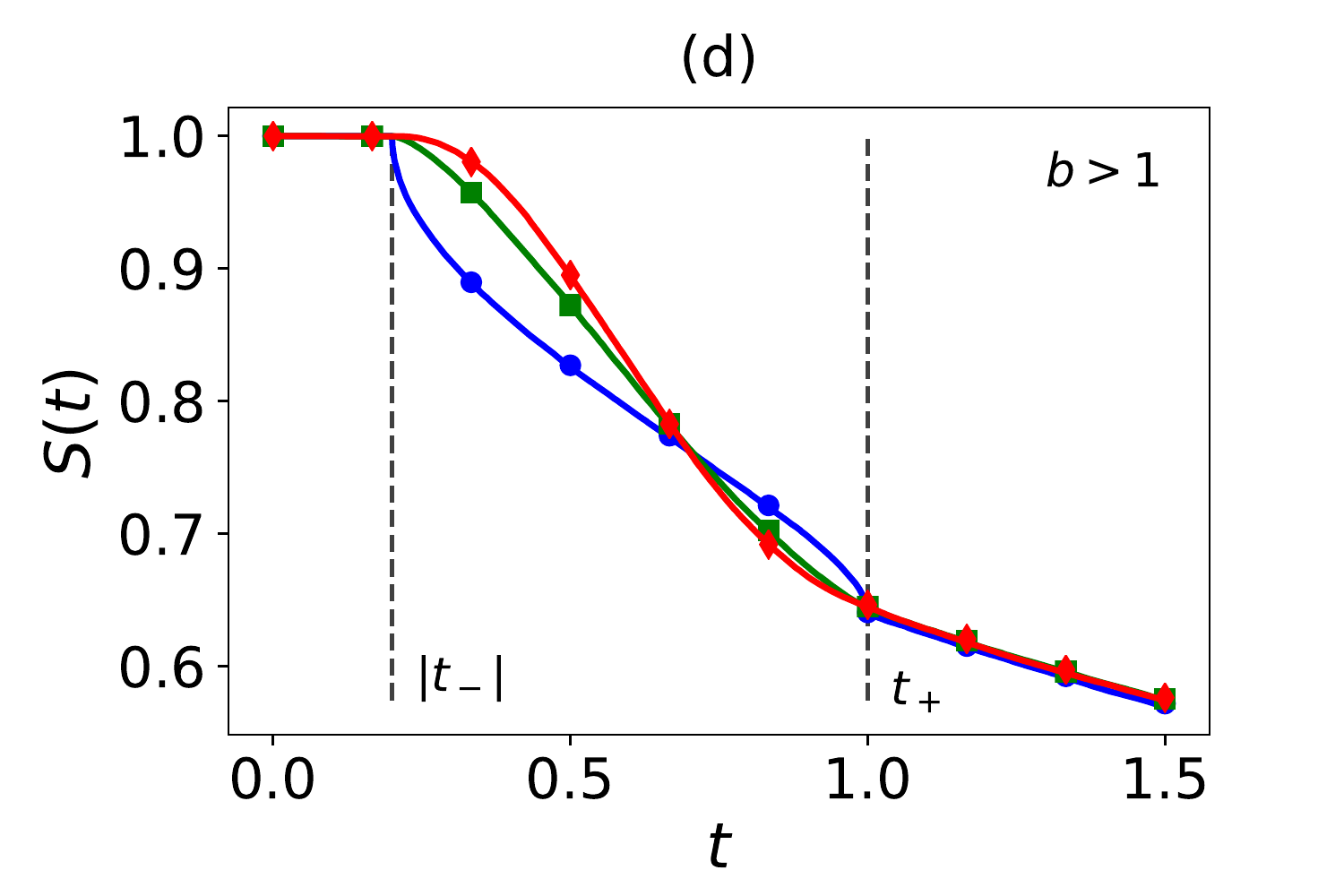}
\caption{In panel (a) and (b) we present, respectively, the first-passage time distribution $F(t)$ density and the survival probability $S(t) = \int_t^{\infty} {\rm d}t' \,F(t')$, in the case where the target is inside the support of the initial distribution $p(x)$ of the RTP in Eq. (\ref{eq:rho}), i.e., with $b<1$. The parameters are: $L=1$, $c=2$, $\gamma=1$, $v=5$, and $\nu=0.5$ (blue curve), $\nu=1.5$ (green curve), $\nu=2.5$ (red curve). We evaluated $F(t)$ from the analytical expressions given in Eqs. (\ref{eq:f0}) and (\ref{eq:f}). In contrast, the integral $S(t)= \int_t^{\infty} {\rm d}t'\, F(t')$ had to be evaluated numerically at few discrete values of $t$ and hence we represent them by symbols in panel (b), even though these results are analytical. We also evaluated $S(t)$ by simulating the corresponding Langevin equation (\ref{eq:RTLangevin}) and the data can be easily obtained for a larger set of discrete times. Hence in panel (b), we represent the numerical Langevin data by solid lines, even though they are numerical. We find an excellent agreement between the analytical prediction and the numerical results. The dashed vertical lines coincide with $t_\pm=\tau_\pm/\gamma$, the times at which particles starting at the edges $\pm c$ reach the target without tumbling. 
Analogously, in panel (c) and (d) we plot $F(t)$ and $S(t)$ for $b>1$, with $L=3$. 
In this case, the support of $F(t)$ is limited from below by $t_-=|\tau_-|/\gamma$, the time at which particles from $+c$ hit the target at $L>c$ without tumbling (see Fig. \ref{Fig_space_two_time}). The second vertical dashed line corresponds to $t_+=\tau_+/\gamma$. Numerical predictions are computed by simulating $N=10^5$ trajectories with a time step $\Delta t=10^{-4}$. Error bars are not visible  on the scale of the plot.}\label{Fig3}
\end{figure*}

\vspace*{0.5cm}
\noindent{\it (ii) The case $b>1$,} the target is outside the support of $p(x)$, i.e.,  $L>c$. In this case, the minimal time $\tau_0$ to reach $L$ is given by $\tau_0=|\tau_-|>0$ (with $\tau_{-} = (c-L)/v<0$). This simply corresponds to the time needed for the particles initially located at $+c$ and moving ballistically with velocity $+v$ (i.e., without tumbling) to reach the target located at $L>c$ -- see the bottom panel of Fig. \ref{Fig_space_two_time}. 
In this case, as for $0<b<1$ discussed above, { we find that the leading contribution to $f(|\tau_\mp|\pm \epsilon)$ is the same as in Eq. \eqref{eq:scaling} with $|\tau_\pm|$ replacing  $\tau_\pm$}. 
In the active regime $0<\nu<1$ (see the blue curve of Fig. \ref{Fig3}c), the divergence of $f(\tau)$ at $|\tau_\mp|\pm\epsilon$ corresponds to fronts of particles, initially at $\pm c$, that hit the target with  constant velocity $\pm v$ without tumbling. In the passive regime $\nu>1$, the MFPT $f(\tau)$ is finite at $|\tau_\mp|$: it displays infinite derivative for $1<\nu<2$ (green curve in Fig. \ref{Fig3}c, or finite one for $\nu\ge2$ (red curve of Fig. \ref{Fig3}c. {As in the case $b<1$, we find that  $f(\tau_++\epsilon)$ is given by Eq. \eqref{eq:ampl2}, while $f(|\tau_-|-\epsilon)=0$ since $|\tau_-|-\epsilon<\tau_0$ is outside the support of $f(\tau)$}. {At long times, the leading behavior of $f(\tau) \sim (1/2 + (\gamma L/v))/\sqrt{2 \pi \tau^{3}}$ -- see Eqs. (\ref{eq:frt0}) and (\ref{eq:ampl}) -- is independent of $c$ and $\nu$. Correspondingly, the curves for different values of  $\nu$ in Fig. \ref{Fig3}c are expected to almost coincide. This is indeed the case, except that, for $\tau > \tau_+$, there are still dependences on $\nu$ (coming from the subleading terms), though they are not visible on the scale of Fig. \ref{Fig3}c. As in the previous case, we have also computed the survival probability $S(t) = \int_t^{\infty} {\rm d}t' F(t')$, which is plotted in Fig. \ref{Fig3}d. The comparison with simulations, once again, is excellent.}


In general, the properties discussed above are found whenever the initial probability density $p(x)$ has qualitative features similar to those discussed here: a finite support, and a transition from \textit{passive} to \textit{active} regime.
For example, we may consider $p(x)$ to be the stationary probability density of an RTP in the more general confining potential $V_q(x)=\mu |x|^q$, where $q>1$. As for the harmonic trap, the corresponding stationary probability density $p_q(x)$ has a finite support $\left(-c_q,c_q\right)$, with $c_q =  [v/(\mu q)]^{1/(q-1)}$ \cite{Confpot}: the modulus of the velocity of the RTPs vanishes at $\pm c_q$ according to the zero velocity condition  $V'_q(\pm c_q)=\pm v$. Moreover, the behavior of $p_q(x)$, in correspondence of points at a distance $\epsilon>0$ from the edges, is given by 

\begin{equation}
    p_q\left(x=\pm c_q \mp \epsilon\right)\sim \epsilon^{\nu(\mu,q)-1},
\end{equation}
where the exponent reads $\nu(\mu,q)=(\mu_c/\mu)^{1/(q-1)}$, with $\mu_c= \left(v/q\right)^{2-q}\left[\gamma/(q(q-1))\right]^{q-1}$.
The \textit{passive} regime is realized for $\nu>1$, corresponding to $\mu<\mu_c$, the \textit{active} one otherwise. 
{As for the case of the harmonic trap, the qualitative properties of the average FPT {near $\tau_\pm$}  depend only on the behavior of $p_q(x)$ near the edges of its support. Namely, except for the potential-dependent prefactor, $f(\tau)$ shows the same scaling as in Eq.~\eqref{eq:scaling}: for $0<b<1$, $f(\tau_\pm-\epsilon)\propto \epsilon^{\nu(\mu,q)-1}$, while for $b>1$, one has  $f(|\tau_\pm|\mp\epsilon)\propto \epsilon^{\nu(\mu,q)-1}$}.

Thus, to summarize, the FPT distribution of an RTP, when averaged over the initial condition with a finite support, exhibits generically two singular points at $\tau = \tau_{\mp}$, corresponding to contributions from the purely ballistic trajectories that originate from the two edges of the supports. They carry the information about the singular behaviour of the initial condition (density) near the two edges. They manifest themselves as singularities in the FPT distribution at $\tau_{\pm}$. This picture is rather generic for an RTP and holds for any initial condition with a finite support.

\section{RTPs with stochastic resetting}\label{sec:RT_reset}

In this section, we study the dynamics of an RTP whose velocity as well as the position are reset at random times as follows. At the initial time, the position $x$ of the particle is randomly
distributed according to the distribution $p(x)$ and it starts with a velocity $\pm v$ with equal probability. The particle then evolves according to Eq. (\ref{eq:RTLangevin}) for a certain random time $\tau$, which is distributed according to an exponential distribution $P(\tau) = r \, {\rm e}^{-r \tau}$, after which the velocity {\it and} the position of the particles are reset instantaneously. The new velocity is set randomly to $\pm v$ with equal probability while the resetting position of the particle is again distributed according to the distribution $p(x)$. The resetting protocol is thus similar to the ``fully-randomized'' protocol introduced in Ref. \cite{Evans_2018} except that here the resetting position is chosen randomly from $p(x)$ at each resetting event. Here we are interested in the case where $p(x)$ is given by Eq.~(\ref{eq:rho}). 

Our main focus here is on the FPT to a target, modeled by an absorbing boundary, at $x=L>0$. Following a renewal approach \cite{Evans_2018,Evans_2020}, we first relate the survival probability $S_r(t)$ with resetting (after averaging over initial positions drawn from a distribution $p(x_0)$) to the averaged survival probability without resetting, i.e., 
\begin{eqnarray}\label{def_S0}
S_0(t) = \int {\rm d}x_0\, S_0(t|x_0)\, p(x_0) \;,
\end{eqnarray}
where $S_0(t|x_0)$ is the survival probability without resetting for a given initial position $x_0$. This relation is best expressed in the Laplace space. We thus define the pair of Laplace transforms 
\begin{eqnarray}
\tilde S_r(s) &=& \int_0^\infty {\rm d}t\, S_r(t) \, {\rm e}^{-st}  \;, \label{LapSr} \\
\; \tilde S_0(s) &=& \int_0^\infty {\rm d}t\, S_0(t) \, {\rm e}^{-st} \label{LapS0} \;.
\end{eqnarray}
Then, the relation between the two reads (see Appendix~\ref{sec:appren})
\begin{equation} \label{Eq_Sr}
    \tilde{S}_r(s)=\frac{\tilde{S}_0(s+r)}{1-r\tilde{S}_0(s+r)}\;.
\end{equation}
Averaging Eq. (\ref{EqS0}) over $p(x_0)$, the Laplace transform $\tilde S_0(s)$ is then given by
\begin{equation}\label{eq:Ss}
    \tilde{S}_0(s)=\frac{1}{s}\left[1+\frac{v\lambda(s)-s-2\gamma}{2\gamma}\int_{-c}^{+c}\mathrm{d}x_0\, {\rm e}^{-\lambda(s)d_L(x)}p(x_0)\right] \;,
\end{equation}
where we recall that $\lambda(s) = \sqrt{s^2+2s\gamma}/v$ and $d_L(x) = |L-x|$.

The FPT distribution $F_r(t)$ is obtained from $S_r(t)$ as ${F_r(t)=-\partial_t{S}_r(t)}$ and the MFPT to the target is thus given by $\langle t_f\rangle=\tilde{S}_r(0)$, yielding from Eq. (\ref{Eq_Sr}),
\begin{equation}\label{eq:tr}
\langle t_f\rangle(r)=\frac{\tilde{S}_0(r)}{1-r\tilde{S}_0(r)} \;,
\end{equation}
where $\tilde S_0(r)$ is given in Eq. (\ref{eq:Ss}). Below, we analyse the behaviour of $\langle t_f\rangle(r)$ as a function of $r$, in the two extreme limits $r \to 0$ and $r \to \infty$, and then for intermediate values of $r$. 

\subsection{MFPT in the limits of small and large $r$}

In order to understand the behavior of $\langle t_f\rangle$ as a function of the resetting rate $r$, we consider separately the two limits $r\rightarrow 0$ and $r\rightarrow\infty.$

\vspace*{0.5cm}
\noindent{\it The limit $r\to 0$.} In this case $\langle t_f \rangle$ is found by Taylor-expanding Eqs. \eqref{eq:Ss} and \eqref{eq:tr} for small $r$, yielding

\begin{equation}\label{eq:tr0}
\langle t_f\rangle(r)=\frac{1}{\sqrt{2\gamma r}}\left[1+\frac{2\gamma}{v}\int_{-c}^{+c}\mathrm{d}x_0\,p(x_0)d_L(x_0)\right]+O(r^0),
\end{equation}
where the convergence of the integral follows from the fact that $p(x_0)$ has a finite first moment.
As in the case of Brownian motion \cite{Reset}, the mean first-passage time diverges as $r^{-1/2}$ for $r\rightarrow 0$, albeit with a different prefactor, as given in Eq.~(\ref{eq:tr0}). 

\vspace*{0.5cm}
\noindent{\it The limit $r\to \infty$.} 
This limit, by contrast, depends crucially on whether $b = L/c<1$ or $b>1$, i.e., whether the target is inside or outside the support. We
consider below the two cases separately. 

\vspace*{0.5cm}

\noindent{\it (i) The case $0<b<1$}. To analyse the large $r$ behavior in Eq. (\ref{eq:tr}), we need to analyse the integral that appears in Eq. (\ref{eq:Ss}), namely
\begin{equation}\label{eq:Itr}
\mathcal{I}(r) = \int_{-c}^{+c}\mathrm{d}x_0\,p(x_0)\exp\left(-\lambda(r) d_L(x_0)\right) \;,
\end{equation}
keeping in mind that $p(x_0)$ has a finite support over the interval $(-c,c)$ and $d_L(x_0) = |L-x_0|$. We start by evaluating the asymptotic behavior of the integral $\mathcal{I}(r)$ in powers of  $r^{-1}$ as $r\rightarrow\infty$. We then make a change of variable $y=(x-L)\lambda(r)$
\begin{equation}\label{eq:intTres1}
\begin{aligned}
\mathcal{I}(r)
&=\frac{1}{\lambda(r)}\int_{-\lambda(r)(L+c)}^{\lambda(r)(c-L)}\mathrm{d}y\,p\left(\frac{y}{\lambda(r)}+L\right){\rm e}^{-|y|}\\
&=\frac{2v}{r}p(L)+O(r^{-2}) \;,
\end{aligned}
\end{equation}
where, using $\lambda = \sqrt{r^2 + 2 \gamma r}/v \sim r/v$ to leading order in $1/r$, we expanded $p(y/\lambda(r)+L)$ and we integrated over $y$. Since $L$ is inside the support of $p(x_0)$, one has $p(L) > 0$. By direct substitution of Eq. \eqref{eq:intTres1} in Eq. \eqref{eq:tr}, we get
\begin{equation}\label{eq:tfrinf}
\langle t_f\rangle(r)=\frac{1}{v\,p(L)}+O(r^{-1}),
\end{equation}
which tells us that $\langle t_f\rangle(r)$ tends to a constant in the large $r$ limit. The fact that $\langle t_f(r)\rangle$ approaches a nonzero constant as $r \to \infty$ is shown in Fig. \ref{fig:tfr}a, where $\langle t_f\rangle(r)$ is plotted for certain representative values of the parameters. This approach
can be either monotonic from above (as in Fig. \ref{fig:tfr}a) or from below. In the latter case, there is a global minimum at some intermediate optimal value $r^*$ as will be discussed in detail later in Section \ref{sec:gen}.


\vspace*{0.5cm}
\noindent{\it (ii) The case $b>1$.} For $L>c$, we can bound the integral in Eq. \eqref{eq:Itr} by
\begin{equation}
\mathcal{I}(r)={\rm e}^{-\lambda(r) L}\int_{-c}^{c}\mathrm{d}x_0\,p(x_0)\,{\rm e}^{\lambda(r)x_0}\le {\rm e}^{-\lambda(r)(L-c)},
\end{equation}
where we have exploited the monotonicity of the exponential. Substituting this inequality in Eqs. (\ref{eq:Ss}) and (\ref{eq:tr}), we obtain a lower bound
for the MFPT 
\begin{equation}\label{eq:expbound}
\langle t_f\rangle(r)\ge \frac{1}{r}\left[\frac{2\gamma \, {\rm e}^{\lambda(r)(L-c)}}{r+2\gamma-v\lambda(r)}-1\right] \;.
\end{equation}
Recalling that $\lambda(r) = \sqrt{r^2 + 2 \gamma r}/v$, we see that the right hand side of this inequality (\ref{eq:expbound})
diverges exponentially as $r\rightarrow \infty$, reflecting the fact that the probability to hit the target outside of the support decreases exponentially at large $r$. Accordingly, for $b>1$, there always exists a finite minimum $r^*$ of the MFPT: this case is shown in Fig. \ref{fig:tfr}b.

\begin{figure}[t!]
\centering
\includegraphics[width = 0.8\linewidth]{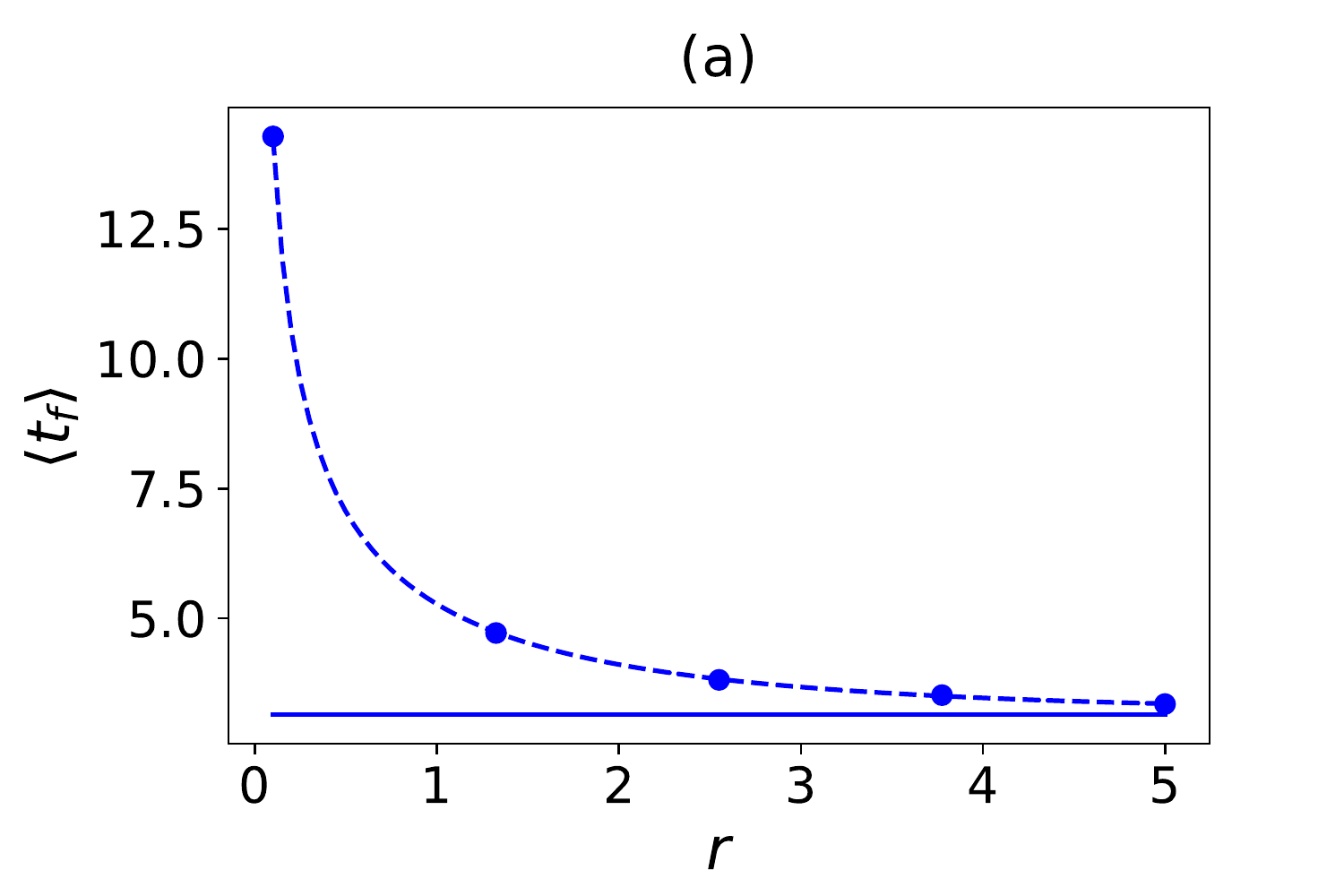}\\
\includegraphics[width = 0.8\linewidth]{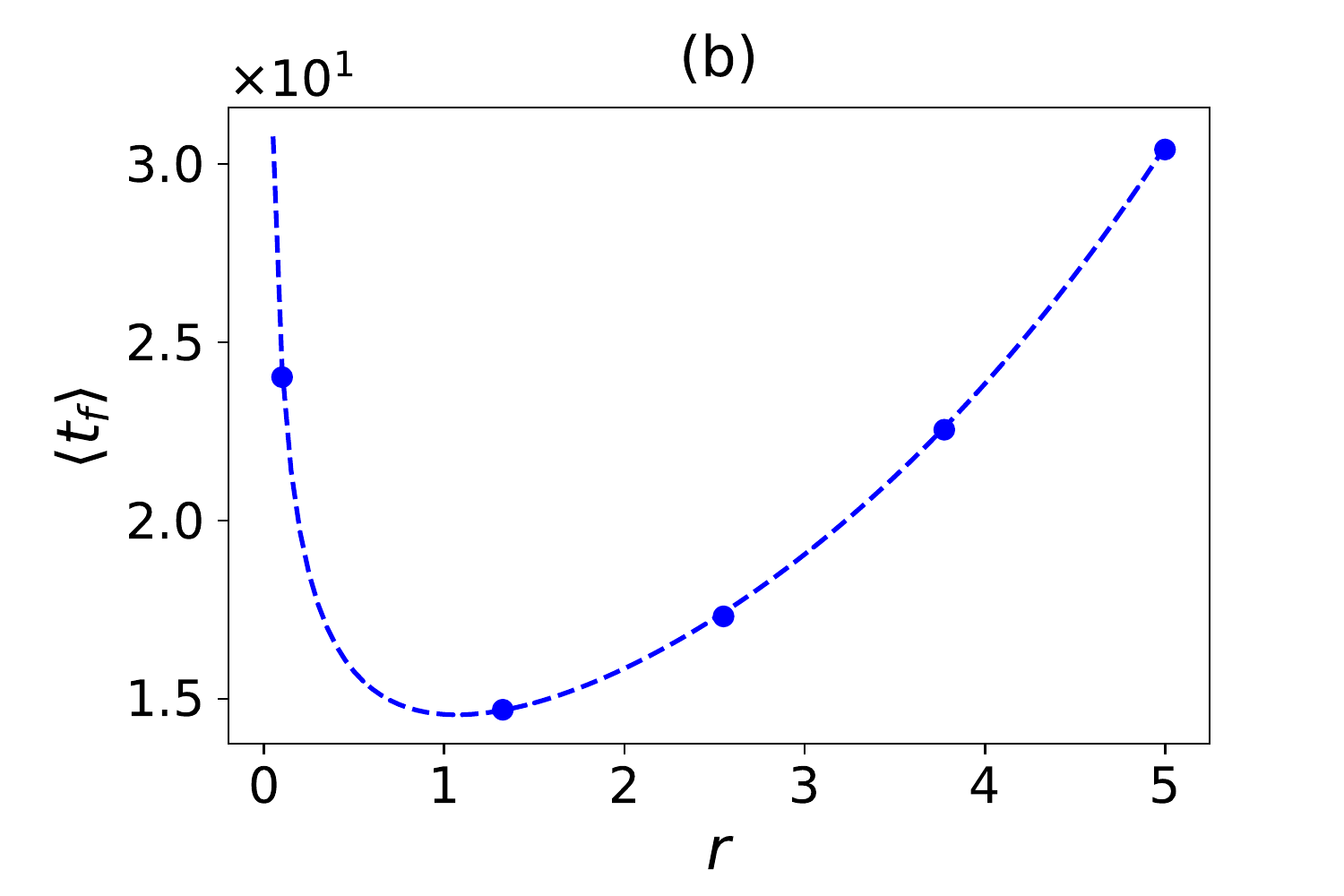}
\caption{Mean first-passage time $\langle t_f\rangle(r)$ in Eq. \eqref{eq:tr} as a function of the resetting rate $r$.
Panel (a) displays $\langle t_f\rangle$ when the target is inside the support of $p(x)$, i.e.,  $b<1$ ($L=2$), and the corresponding asymptotic value $(v p(L))^{-1}$ is denoted by the horizontal line. In panel (b), the target is outside the support, i.e., $b>1$ ($L=3$). The values of the relevant parameters are $v=1.5$, $\gamma=2.5$, $c=2.5$, and $\nu=0.5$. 
In both figures, the dashed line represents the analytical expression in Eq. \eqref{eq:tr}, while dots refer to the corresponding numerical results, computed by simulating $N=4\times 10^4$ trajectories with time step $\Delta t=10^{-3}$. Error bars, not always visible on the scale of the plot, are computed by standard deviation over the ensemble average.}\label{fig:tfr}
\end{figure}

\subsection{MFPT for intermediate values of $r$}

To analyse Eqs. (\ref{eq:Ss}) and (\ref{eq:tr}) for intermediate values of $r$, it turns out to be convenient to use 
the dimensionless variables
\begin{eqnarray} \label{def_X}
X =  L\lambda(r) \;, \; a = \frac{\gamma L}{v} \;, \; b = \frac{L}{c} \;,
\end{eqnarray} 
where we recall that $\lambda(r) = \sqrt{r^2 +2 \gamma r}/v$ and $L$ and $c$ denote respectively the location of the target and the right edge of the support of the initial distribution 
$p(x_0) = c^{-1} \rho(x_0\,c^{-1})$, with $\rho(z)$ given in Eq. (\ref{eq:rho}). We define the dimensionless MFPT $\langle \tau_f \rangle =\gamma \langle t_f \rangle/(8a^2)$, which can be expressed, using Eqs. (\ref{eq:Ss}) and (\ref{eq:tr}) as
\begin{equation}\label{eq:tauf}
\begin{aligned}
&\langle\tau_f\rangle =  \frac{\gamma \langle t_f\rangle}{8a^2}=\frac{1}{4a X^2}\\
&\times\left[(2a+\sqrt{4a^2+X^2})\left(\frac{1}{\rho_r(X,b)}-1\right)+\frac{X}{\rho_r(X,b)}\right],
\end{aligned}
\end{equation}
where we denote 
\begin{eqnarray}\label{eq_rhor}
\rho_r(X,b)=\int_{-1}^{+1}\mathrm{d}z\,\rho(z) \,{\rm e}^{-X\,l(z)} \;,
\end{eqnarray}
with $l(z) =   |{z}/{b}-1|$ and $\rho(z)$ given in Eq. (\ref{eq:rho}). Thus Eq. (\ref{eq:tauf}) gives the 
rescaled MFPT as a function of the resetting rate $r$ through the variable $X = L \lambda(r) = L \sqrt{r^2 +2 \gamma r}/v$ for 
fixed values of the two parameters $a$, $b$ as well as the initial rescaled density $\rho(z)$ with a finite support with edges at $z =\pm 1$. 

Below we analyse this Eq.~(\ref{eq:tauf}) as a function of $X$, and hence of $r$, keeping $a$, $b$ and $\rho(z)$ fixed. It turns out to be convenient 
to discuss the simpler diffusive limit first, where $\gamma \to \infty$, $v \to \infty$ with fixed $v^2/(2 \gamma) = D$ [see Eq.~(\ref{diff_lim})]. In this limit $a   = \gamma L/v = \sqrt{\gamma/(2D)}L \to \infty$, while $b$ and $\rho(z)$ are kept fixed. This is done below in Section \ref{sec:diff}, followed by the analysis of the generic case in Section \ref{sec:gen}.

\subsubsection{The diffusive limit of the RTP} \label{sec:diff}

In the diffusive limit, the parameter $a \to \infty$, as discussed above, while we keep $b$ and $\rho(z)$ fixed. Furthermore, we recall that the initial scaled distribution, $\rho(z) = N(\nu)\,(1-z^2)^{\nu-1}$ for $-1 < z <1$ with $N(\nu)$ given in Eq. (\ref{Nu}), is characterised by a single parameter $\nu > 0 $. Thus, in this limit, the rescaled MFPT in Eq. (\ref{eq:tauf}) reduces to  
\begin{equation}\label{eq:tBM}
    \langle\tau_f\rangle=\frac{1}{X^2}\left[\frac{1}{\rho_r(X,b)}-1\right] \;,
\end{equation}
with $\rho_r(X,b)$ given in Eq. (\ref{eq_rhor}). We analyse $\langle\tau_f\rangle$ in Eq.~(\ref{eq:tBM}) as a function of $X= L\sqrt{{r}/{D}}$ for two fixed parameters $b$ and $\nu>0$. In fact, for $b>1$, the scaled $\langle\tau_f\rangle$ has always a minimum (as a function of $r$) 
at some optimal value $r^*$ [see Fig. \ref{fig:tfr}b], irrespective of the parameter $\nu$. In contrast, more interesting behavior emerges, as shown below, for the complementary case $0<b<1$, when the target is inside the support of the initial distribution. Hence, below, we focus on $0<b<1$, considering various values of the parameter~$\nu$.

We first focus on the two extreme limits $X \ll 1$ and $X \to \infty$. To understand their physical significance, it is useful to rewrite $X = L \sqrt{r/D} = \sqrt{\tau_d/\tau_r}$
where $\tau_d=L^2/D$ and $\tau_r = 1/r$. Note that $\tau_d$ is the typical time to cover a distance $L$ purely by diffusion and $\tau_r$ is the typical time between two successive resettings. When $X \ll 1$, i.e., for $r\to 0$, the resetting is rare compared to diffusion. From Eq. (\ref{eq_rhor}), expanding for small $X$ and using the normalisation condition $\int_{-1}^{+1}  {\rm d}z \, \rho(z) = 1$, we get
\begin{eqnarray} \label{rhor_smallX}
\rho_r(X,b) = 1 - X \int_{-1}^{+1}  {\rm d}z \, \rho(z) l(z) + O(X^2) \;,
\end{eqnarray}
where $l(z) = |z/b-1|$. Substituting this result in Eq.~(\ref{eq:tBM}), we find 
\begin{equation}\label{eq:tBM2}
    \langle\tau_f\rangle = \frac{\int_{-1}^{+1}  {\rm d}z  \, \rho(z) l(z)  }{X} + O(X^0) \quad, \quad X \to 0 \;.
\end{equation}    
In the opposite limit where $X \to \infty$ (when resetting is more frequent than diffusion), we first analyse $\rho_r(X,b)$ in Eq. (\ref{eq_rhor}). Performing the change of variable $X(z/b-1) = y$, we get
\begin{eqnarray} \label{change}
\rho_r(X,b) = \frac{b}{X} \int_{-X(1+\frac{1}{b})}^{X(\frac{1}{b}-1)} {\rm d}y \, \rho\left( b + \frac{by}{X}\right)\, {\rm e}^{-|y|} \;.
\end{eqnarray}  
Note that the upper limit in the integral approaches $+ \infty$ as $X \to \infty$ (since $0<b<1$) while the lower limit approaches $-\infty$. Expanding for large $X$, we then get
\begin{eqnarray}\label{rhor_largeX}
\rho_r(X,b) = \frac{2b\,\rho(b)}{X}\,  + O (X^{-3})  \;.
\end{eqnarray}
Consequently, $\langle \tau_f \rangle$ in Eq. (\ref{eq:tBM}) behaves as
\begin{eqnarray} \label{eq:tBM3}
\langle \tau_f \rangle = \frac{1}{2 b \, \rho(b)} \frac{1}{X}  + O(X^{-3}) \;.
\end{eqnarray}
Thus, from Eqs. (\ref{eq:tBM2}) and (\ref{eq:tBM3}), we see that $\langle \tau_f \rangle$ diverges as $X \to 0$ as $\langle \tau_f \rangle \propto 1/X$ and it decays very slowly still as $\langle \tau_f \rangle \propto 1/X$, for large $X$. 

The interesting question is then: how does $\langle \tau_f \rangle$ behave for intermediate values of $X$, between these two extreme limits? For instance, does $\langle \tau_f \rangle$ decrease monotonically upon increasing $X$ or is there any possibility of a non-monotonic behavior? Indeed, it turns out that this monotonicity depends on both parameters $\nu$ and $b$. By evaluating $\langle \tau_f\rangle$ numerically from Eq. (\ref{eq:tBM}), we generically find two types of behavior, depending on $\nu$ and $b$: (i) the function $\langle \tau_f\rangle$ decreases monotonically upon increasing $X$, implying that $r \to \infty$ is the optimal resetting rate and (ii) the function $\langle \tau_f\rangle$ develops an additional local minimum at $X = X_{\min}$ -- {we call it a ``kink'' in the following}. However, this minimum is ``{metastable}'' in the sense that $\tau_{\min} = \langle \tau_f \rangle(X=X_{\min})$ is larger than the true global minimum which occurs always at $X \to \infty$. A similar 
metastable behavior was also noticed in the theory and experiments of pure Brownian diffusion, but starting only from the Gaussian initial distribution with a finite width $\sigma$~\cite{Ciliberto, Besga,Faisant_2021}. In our study here, the initial distribution (corresponding to the stationary distribution of an RTP in a harmonic trap), which also has a finite width but an additional parameter $\nu$ which can be tuned to generate a family of shapes of the initial distribution. This leads to a richer phase diagram in the two-parameter plane as discussed below.

\begin{figure}[t]
\includegraphics[width = \linewidth]{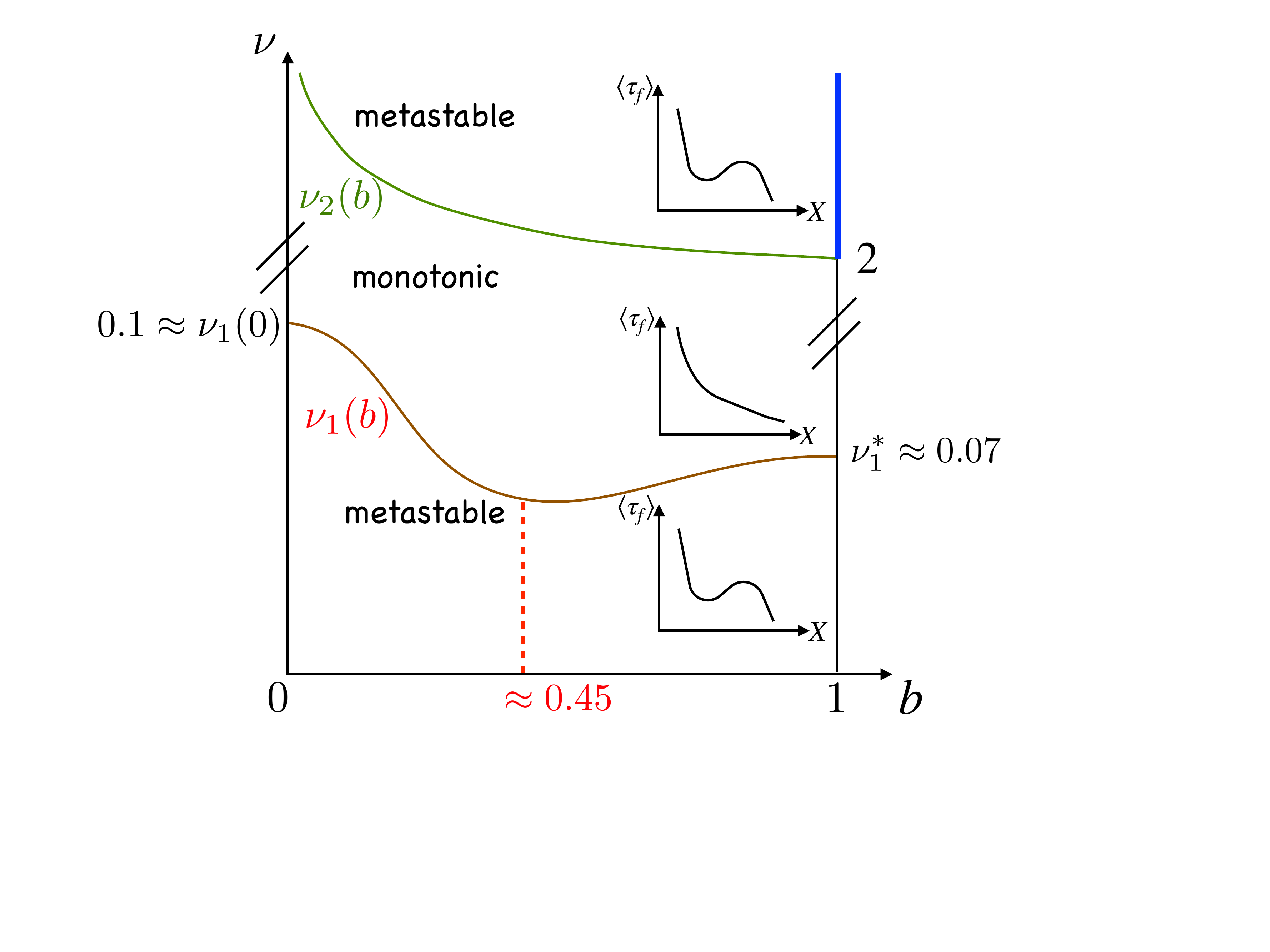}
\caption{Sketch of the phase diagram for the simpler diffusive limit in the $(b,\nu)$ plane (see the text for details).}\label{Fig_phdiag}
\end{figure}
Our findings are summarised in the ``phase diagram'' in the $(b, \nu)$ plane shown in Fig. \ref{Fig_phdiag}. In this plane, there are two lines $\nu_1(b)$ and $\nu_2(b)$ ($> \nu_1(b)$), for $0\leq b<1$. For $\nu < \nu_1(b)$, there is a metastable minimum ({i.e., a kink}) at $X = X_{\min}$ and the rescaled MFPT $\langle \tau_f\rangle$ in Eq. (\ref{eq:tBM}), as a function of $X$, has a local (but not global) minimum
at some finite $X$ (see Fig. \ref{Fig_phdiag}). We call this phase ``metastable''. For $\nu_1(b)<\nu<\nu_2(b)$, the rescaled MFPT is a monotonically decreasing function of $X$ -- we call this phase ``monotonic''. When $\nu$ exceeds $\nu_2(b)$, a kink develops again in the $\langle \tau_f \rangle$ vs. $X$  curve, indicating the re-appearance of the metastable phase. Thus, we find a novel re-entrance ``phase transition'' across the lines $\nu_1(b)$ and $\nu_2(b)$. {In the limit $b \to 0$, we find that $\nu_2(b \to 0) \to \infty$ while $\nu_1(b \to 0) = \nu_1(0) \approx 0.1$}. In the other limit $b \to 1$, we find that {$\nu_1(b\to 1) = \nu_1^* \approx 0.07$, while $\nu_2(b \to 1) = 2$ (see below). Our numerical simulations indicate that the curve $\nu_1(b)$ is nonmonotonic as $b$ approaches the value $b=1$ (see Fig. \ref{Fig_phdiag}). In addition, it turns out that the behavior exactly at $b=1$ is different from the limit $b \to 1^-$ in the following sense. Indeed, exactly at $b=1$, there are only two phases (instead of three in the limit $b {\to} 1^-$), depending on whether $\nu < \nu_2(b=1)=2$ or $\nu > \nu_2(b=1)=2$. For 
$0<\nu < \nu_2(b=1)=2$, the scaled MFPT $\langle \tau_f\rangle$ is a monotonically decreasing function of $X$, thus indicating that $\nu_1(b=1) = 0$. This implies that $\nu_1(b \to 1) \approx 0.07 \neq \nu_1(b=1) = 0$. In addition, the phase for $\nu > \nu_2(b=1)=2$ is slightly different from the metastable phase for $b<1$. Indeed, it turns out that for $b=1$ and $\nu>\nu_2(b=1)=2$, the scaled MFPT $\langle \tau_f\rangle$ has a single global minimum at $X=X_{\min} < \infty$, which thus is not metastable. This phase is shown by the solid vertical blue line in the phase diagram (see Fig. \ref{Fig_phdiag}). As one crosses the phases boundaries $\nu_1(b)$ and $\nu_2(b)$, for fixed $b$, upon increasing $\nu$, the transition from the metastable to the monotonic phases is somewhat similar to the ``spinodal'' transition that happens in thermodynamics systems.}

This reentrance transition, as the parameter $\nu$ increases, can be qualitatively understood by the following argument. The parameter $\nu$ controls the location of the peak of the initial distribution. For example, when $0< \nu < 1$, the peak of the initial distribution occurs close to $z=1$ (and symmetrically at $z=-1$), i.e., more particles are concentrated, in the initial condition, at the edges of the support. We recall that the target is located at a fixed scaled distance $b=L/c<1$ to the right of the origin. Thus for small $\nu$, the peak of the initial distribution at $z=1$ is to the right of the target and well separated from it. A similar situation arose in Refs. \cite{Ciliberto, Besga,Faisant_2021},  
where the initial distribution was Gaussian, centered at the origin with a finite width $\sigma$. There, it was shown that if the target and the peak of the Gaussian initial distribution are well separated, indeed one finds a metastable state with a kink, while the true minimum still occurs at $X_{\min}= \infty$. This nonmonotonic 
decay of $\langle \tau_f \rangle$ vs. $X$ is shown schematically in the inset of the lower metastable phase in Fig. \ref{Fig_phdiag}. In the opposite limit $\nu \gg 1$, the peak of the initial distribution will be concentrated around $z=0$, i.e., on the left side of the target at $b$ and again clearly separated from it. Since the diffusion is symmetric, this is qualitatively similar to the case when the initial peak was to the right of the target. Thus, one would again expect a metastable state with a kink in the curve $\langle \tau_f \rangle$ vs. $X$, as shown schematically in the inset of the upper metastable phase in Fig. \ref{Fig_phdiag}. For the intermediate values of the parameter $\nu$, the peak of the initial distribution is rather close to the target and, hence, following the argument of Refs. \cite{Ciliberto, Besga,Faisant_2021}, one would expect a monotonic decay of $\langle \tau_f \rangle$ vs $X$, as shown schematically in the inset of the middle phase in Fig. \ref{Fig_phdiag}. However, this qualitative argument does not provide a detailed location of the phase boundaries $\nu_1(b)$ and $\nu_2(b)$, for which one needs to analyse Eq. (\ref{eq:tBM}) in more details (which we did numerically). While it is difficult to extract the analytical expressions of the two curves $\nu_1(b)$ and $\nu_2(b)$, we can estimate them numerically for generic $b$. However, the two limiting cases $b \to 0$ and $b \to 1^-$ can be studied analytically, as we now show. \\

\vspace*{0.5cm}
\noindent{\it The case $b \to 0$}. This corresponds to the target being located at the origin. From the expression of $\rho_r(X,b)$ in Eq. (\ref{eq_rhor}), which explicitly reads
\begin{eqnarray}
\rho_r(X,b)=\int_{-1}^{+1}\mathrm{d}z\,\rho(z) \,{\rm e}^{-X\,\left|\frac{z}{b}-1\right|} \;, \label{rhor_1}
\end{eqnarray}
one sees that in the limit $b \to 0$, it becomes only a function of $X/b$, 
\begin{eqnarray}\label{rhorb_smallb}
\rho_r(X,b) &\approx & \tilde \rho_r \left( Y = \frac{X}{b}\right) \;, \nonumber \\
{\rm where}\;\; \tilde \rho_r(y) &=& \int_{-1}^{1} {\rm d}z\, \rho(z)\,{\rm e}^{-Y |z|} \;.
\end{eqnarray}
Plugging this result in Eq. (\ref{eq:tBM}), we see that the scaled MFPT can be expressed in a scaling form when $b \to 0$
as 
\begin{eqnarray}
\langle \tau_f\rangle &\approx& \frac{1}{b^2} \tilde{W}\left( \frac{X}{b}\right) \quad {\rm with} \nonumber \\
 \tilde W(Y) &=& \frac{1}{Y^2} \left[\frac{1}{2\int_0^1 {\rm d}z \, \rho(z) \, {\rm e}^{-Yz}} -1\right] \;, \label{tau_smallb}
\end{eqnarray}
where we used $\rho(z) = \rho(-z)$ (see Eq. (\ref{eq:rho})). By plotting the scaling function $\tilde W(Y)$ vs $Y$, we see two types of behaviours depending
on the value of $\nu$ that characterizes $\rho(z)$ given in Eq. (\ref{eq:rho}). For $\nu<\nu_1(0)$, the function $\tilde W(Y)$ exhibits a metastable behavior, while
for $\nu > \nu_1(0)$, the curve $\tilde W(Y)$ vs. $Y$ is a monotonically decreasing function (see Fig. \ref{Phdiag_beq0}). To determine this critical point $\nu_1(0)$, we notice that when the
metastable minimum disappears, both the first and the second derivative of $\tilde W(Y)$ vanish at the value $Y=Y_c$, thus making it a point of inflection (see Fig. \ref{Phdiag_beq0}). In this sense, this is a ``spinodal phase transition''. Setting $\tilde W'(Y_c)=0$ and $\tilde W''(Y_c)=0$, we have two equations for the two unknowns $Y_c$ and $\nu_1(0)$. This determines $\nu_1(0)$ and numerically we find $\nu_1(0) \approx 0.1$. 
 In this case, there is no reentrance phase transition at $\nu = \nu_2(0)$, since $\nu_2(0) \to \infty$. As opposed to $b>0$, where the right peak of the initial distribution can cross the target from right to left as $\nu$ increases (leading to the second metastable phase for $\nu > \nu_2(b)$), in the limit $b \to 0$, this crossing cannot happen indicating that for any $\nu>\nu_1(0)$, the phase must be monotonic. This leads to the divergence of $\nu_2(b)$ as $b \to 0$. 

\begin{figure}[t]
\includegraphics[width = \linewidth]{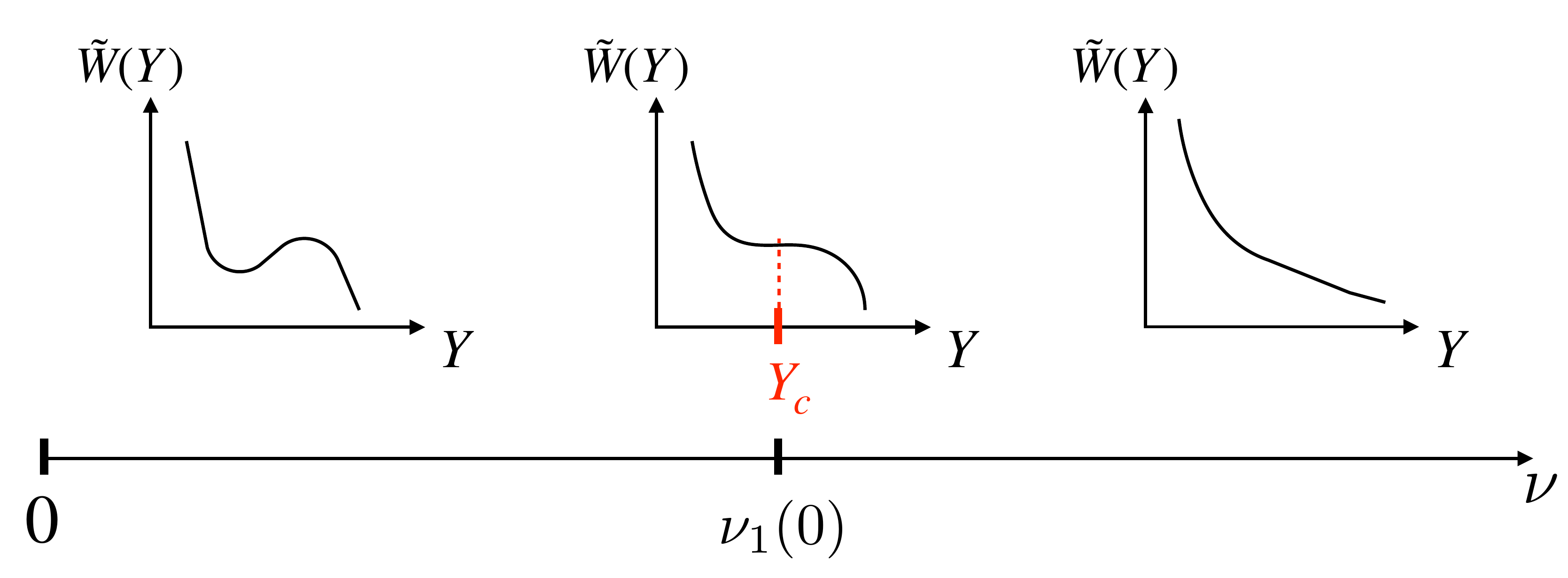}
\caption{Sketch of $\tilde W(Y)$ in Eq. (\ref{tau_smallb}) as a function of $Y$ as $\nu$ increases across the critical value $\nu_1(0)$.}\label{Phdiag_beq0}
\end{figure}

\vspace*{0.5cm}
{\noindent{\it The case $b \to 1^-$}. In this case, the numerical study of $\langle \tau_f \rangle$ in the metastable phase $\nu<\nu_1(b)$ shows that the location of its local minimum $X_{\min}$ diverges as $X_{\min} \propto 1/(1-b)$. It is thus natural to study $\langle \tau_f \rangle$ in Eq. (\ref{eq:tBM}), and hence $\rho_r(X,b)$ in Eq. (\ref{eq_rhor}), in the scaling limit $b \to 1$, $X \to \infty$ but keeping the product $Z = X(1-b)$ fixed. To study $\rho_r(X,b)$ in this limit, it is convenient to start from the expression given in Eq. (\ref{change}), obtained from the original expression in Eq.~(\ref{eq_rhor}) after a simple change of variable. After straightforward manipulations, one finds that, in this scaling limit, $\rho_r(X,b)$ takes the scaling form
\begin{eqnarray}\label{rho_scaling_beq1}
\rho_r(X,b) \approx (1-b)^\nu \, \bar{\rho}_{r}(Z = (1-b)X) \;,
\end{eqnarray}
where the the scaling function $\bar{\rho}_{r}(Z)$ is given by
\begin{eqnarray}\label{rho_bar}
\bar{\rho}_{r}(Z) = \frac{2^{\nu-1}\,N(\nu)}{Z^\nu} \int_{-\infty}^Z {\rm d}y \, (Z-y)^{\nu-1} {\rm e}^{-|y|} \;,
\end{eqnarray}
with $N(\nu) =  \Gamma\left(\nu+1/2\right)/(\sqrt{\pi}\,\Gamma\left(\nu\right))$. By substituting this scaling form (\ref{rho_scaling_beq1})-(\ref{rho_bar}) in Eq. (\ref{eq:tBM}), one finds that in this scaling limit $\langle \tau_f \rangle$ reads
\begin{eqnarray}\label{tau_b1}
\langle \tau_f \rangle \approx (1-b)^{2-\nu} \overline{W}(Z)
\end{eqnarray}
where
\begin{eqnarray}\label{Wbar}
\overline{W}(Z) = \frac{2^{1-\nu}}{N(\nu)} \frac{Z^{\nu-2}}{\int_{-\infty}^Z {\rm d}y \, (Z-y)^{\nu-1} {\rm e}^{-|y|}} \;.
\end{eqnarray}
One can then analyse the function $\overline{W}(Z)$ in Eq. (\ref{Wbar}) exactly as we did before for the function $\tilde W(y)$ in Eq. (\ref{tau_smallb}). By varying $\nu$, we actually
find a behavior qualitatively similar to the one depicted in Fig. \ref{Phdiag_beq0}, with $\nu_1(0)$ replaced by a different value $\nu_1^* = \lim_{b \to 1}\nu_1(b) \approx 0.07$. Here also, this critical value separates a phase, for $\nu < \nu_1^*$, where $\overline W(Z)$ exhibits a nonmonotonic behavior with a local minimum at $Z = Z_{\min}$ from a phase, for $\nu > \nu_1^*$, where $\overline{W}(Z)$ is monotonically decreasing. Exactly at $\nu = \nu_1^*$, the scaling function $\overline{W}(Z)$ exhibits an inflection point at some point $Z=Z_c$ (as shown for $\tilde W(Y)$ in Fig. \ref{Phdiag_beq0}). 
}

\vspace*{0.5cm}
\noindent{\it The case $b  =1$.} This special value is singular and needs to be treated separately. In this case, the integral for $\rho_r(X,b=1)$ in Eq. (\ref{eq_rhor}) can be computed explicitly (see Appendix \ref{app_bessel}) and it reads
\begin{equation}
\rho_r(X,b=1) = \Gamma\left(\nu+\frac{1}{2}\right) \left(\frac{X}{2}\right)^{\frac{1}{2}-\nu}I_{\nu-\frac{1}{2}}(X)\,{\rm e}^{-X}\;, \label{rho_bessel}
\end{equation}
where $I_{\alpha}(x)$ is the modified Bessel function of index $\alpha$. Substituting this result in Eq.~(\ref{eq:tBM}), we get an explicit expression of $\langle \tau_f \rangle$ as a function of $X$. Let us first examine the $X \to 0$ behavior. In this limit, it is easy to see that 
\begin{eqnarray}\label{tauf_smallX_b1}
\langle \tau_f \rangle \approx \frac{1}{X} \quad, \quad {\rm as}\quad X \to 0 \;.
\end{eqnarray}
Thus $\langle \tau_f\rangle$ diverges as $X \to 0$ with an amplitude which is independent of $\nu$. We next consider the opposite limit $X \to \infty$. Taking this limit in Eqs. (\ref{rho_bessel}) and (\ref{eq:tBM}), we find 
\begin{eqnarray}\label{tauf_largeX}
\langle \tau_f \rangle \approx \frac{\sqrt{2\pi}}{2^{\nu-\frac{1}{2}} \Gamma(\nu+1/2)} X^{\nu-2} \;, \; {\rm as}\; X \to \infty \;.
\end{eqnarray}
Accordingly $\langle \tau_f \rangle$ diverges as $X \to \infty$ for $\nu > 2$. This indicates that $\langle \tau_f \rangle $ is a nonmonotonic function of $X$ for $\nu>2$. By plotting this function, one can see indeed that it has a unique minimum at $X=X_{\min}$ for all $\nu > 2$. In contrast, for $\nu<2$, the expression in Eq. (\ref{tauf_largeX}) indicates that $\langle \tau_f \rangle$ decays to $0$ as $X \to \infty$, hinting that the function $\langle \tau_f \rangle$ may be monotonic for any $X>0$. By plotting $\langle \tau_f \rangle$ for $\nu < 2$, we see that it is indeed a monotonically decreasing function of $X$. This last fact shows that $\nu_1(b =1) =0$ and $\nu_2(b=1)=2$.  {As discussed above, we thus see that the curve $\nu_1(b)$ is discontinuous since $\nu_1(b \to 1) \approx 0.07 >  \nu_1(b=1) = 0$.} \\

\subsubsection{The more general RTP case}\label{sec:gen}

We now consider the more general RTP case, where the parameter $a = \gamma L/b$ is now finite (recall that in the diffusive limit discussed above, the parameter $a \to \infty$). 
For finite $a$, we need to investigate Eqs. (\ref{eq:tauf}) and (\ref{eq_rhor}) and plot $\langle \tau_f \rangle$ as a function of $X$ for fixed parameters $a$, $b$ and $\nu$. Thus, compared to the previously discussed diffusive case, we have an additional parameter $a$ here. It turns out that the finiteness of the parameter $a$ induces interesting changes on the $\langle \tau_f \rangle$ vs $X$ curve for fixed parameters $a, b$ and $\nu$. We recall that we only consider the case $b=L/c < 1$ such that the target is inside the support of the initial distribution. In the case $b>1$, we have seen that there is always a true global minimum in the $\langle \tau_f \rangle$ vs $X$ at a finite value $X=X_c$ and there is no metastable phase. In contrast, for $b<1$, both metastable and monotonic phases may appear, as we have seen in the limit $a \to \infty$. Hence, here, we focus on $b<1$ but with $a$ finite. 

It turns out that in this case of finite $a$ and $b<1$, the $\langle \tau_f \rangle$ vs $X$ develops additional features as summarised in Fig. \ref{Fig:reset}. We see from Fig. \ref{Fig:reset}a that  $\langle \tau_f \rangle$ always approaches a constant asymptotically as $X \to \infty$ (see the discussion in Eq.~(\ref{eq:tfrinf}}) and below). However, the approach to this asymptotic constant may be either monotonic or nonmonotonic, depending on the parameter values. The three representative cases are shown in Fig.~\ref{Fig:reset}a where we plotted $\langle \tau_f \rangle$ vs $X$ for $a=0.5$, $b=0.95$ and three different values of $\nu = 0.2, 1.0$ and $\nu = 1.5$. For $\nu = 0.2$ (the green line), the curve develops a global minimum at some value of $X_{\min}$ and then increases before finally approaching the asymptotic constant from above. For $\nu=1.5$ (the blue line), this curve again has a global minimum (though a shallow one) after which it increases monotonically to approach the asymptotic constant from below. For the intermediate value $\nu  =1.0$ (the red line), the curve approaches the asymptotic constant purely monotonically from the beginning. Thus, this is somewhat different from the diffusive $a \to \infty$ limit. Here, as the parameter $\nu$ increases, we again have a reentrance transition but from a ``true minimum'' to another ``true minimum'' phase, separated by a monotonic phase in-between, where the minimum occurs at $X \to \infty$. The location $X_{\min}$ of  the global minimum is plotted as a function of increasing $\nu$ in Fig. \ref{Fig:reset} for fixed $a=0.5$ and $b=0.95$. We see that for $\nu<\nu_1(a,b)$, $X_{\min}$ is finite and increases with $\nu$. When $\nu_1(a,b)<\nu<\nu_2(a,b)$, the location of the minimum $X_{\min}$ jumps to $+ \infty$ (shown by the shaded region). For $\nu > \nu_2(a,b)$, the location $X_{\min}$ again becomes finite and increases further upon increasing $\nu$.   
For the RTP, this is the analogue of the reentrance phase transition discussed before in the diffusive limit (see Fig. \ref{Fig_phdiag}).

\begin{figure}[t!]
\centering
\includegraphics[width = 0.8 \linewidth]{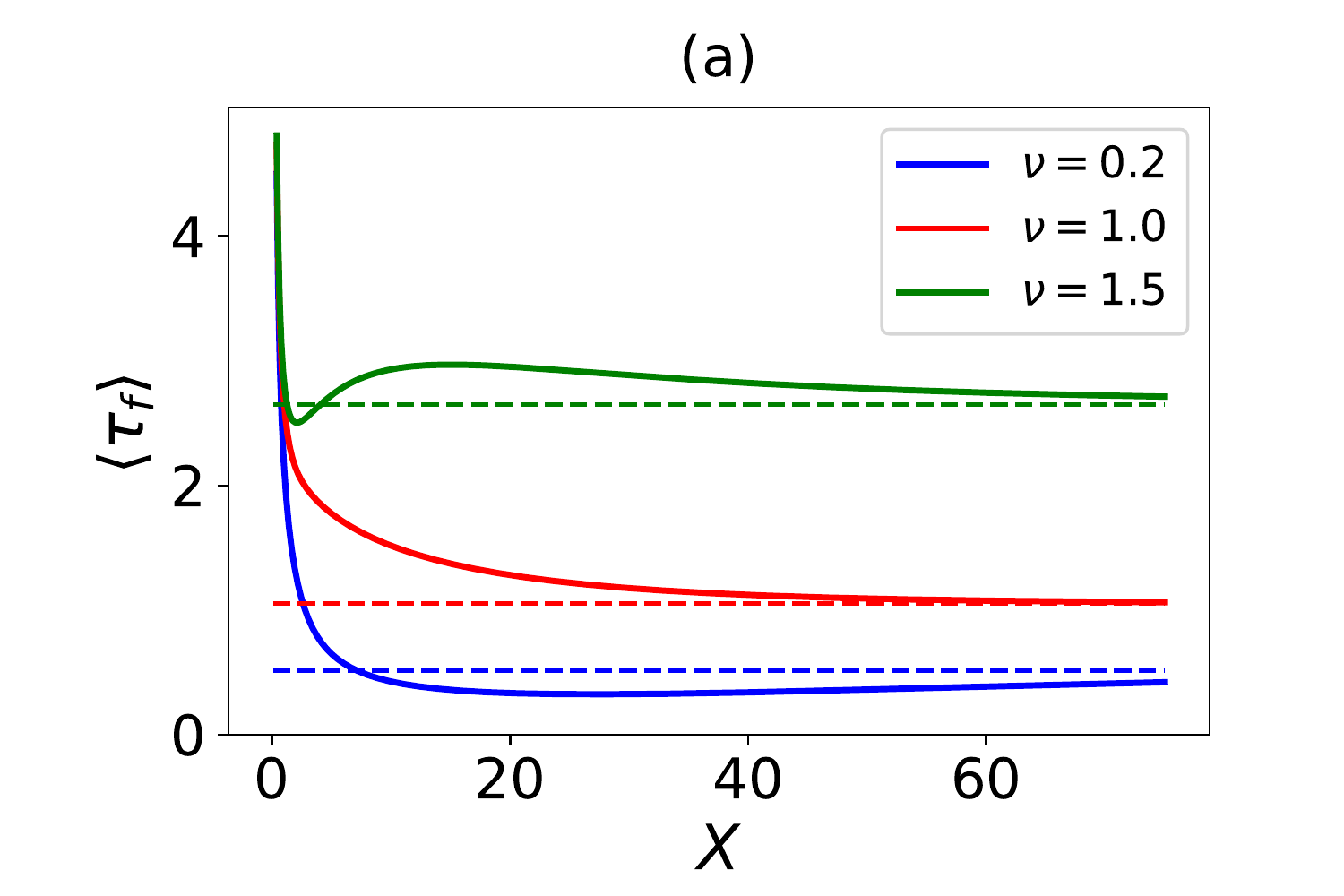}\\
\includegraphics[width = 0.8 \linewidth]{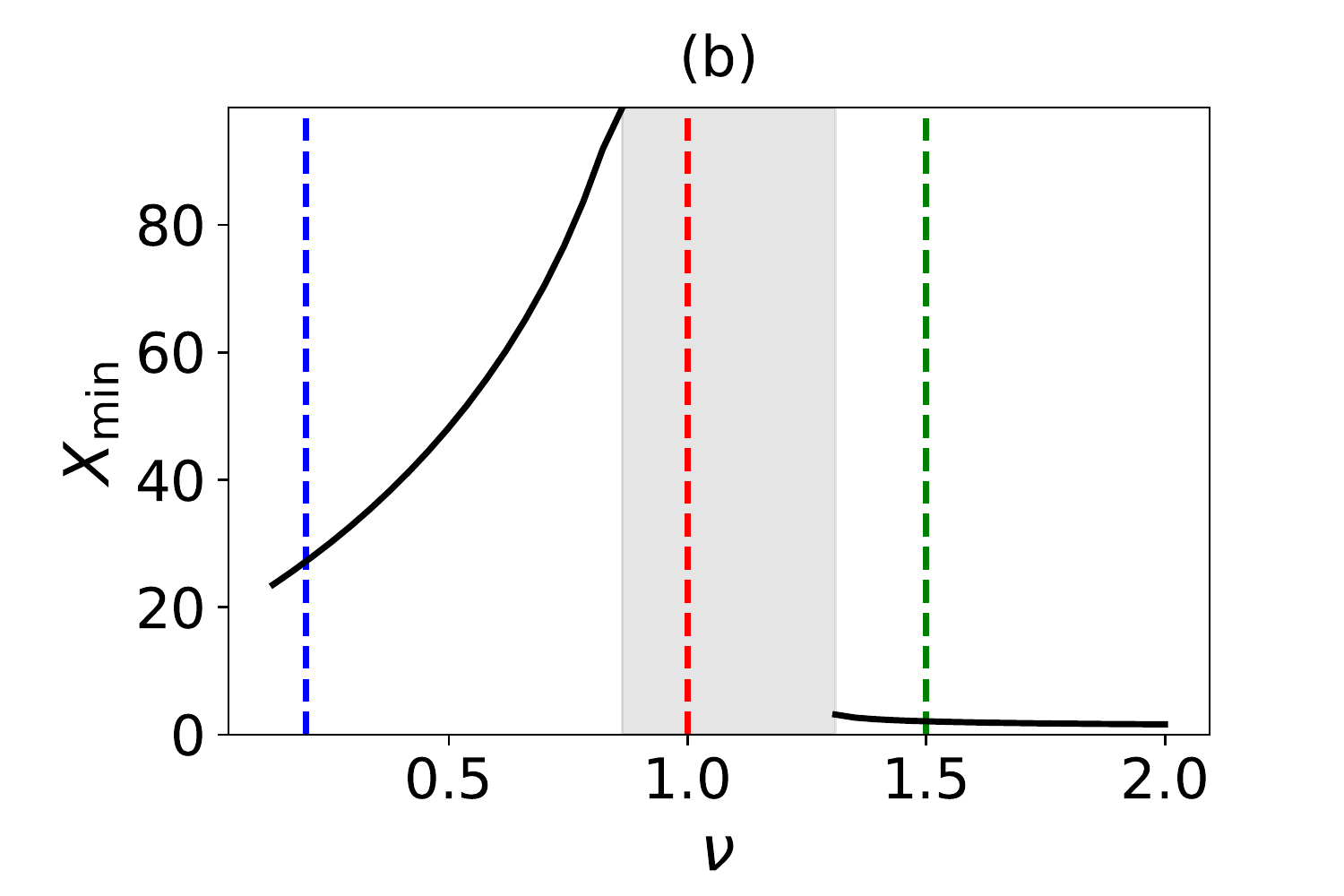}
\caption{In panel (a), the dimensionless mean first-passage time $\langle \tau_f\rangle$ in Eq. \eqref{eq:tauf} is plotted as a function of $X$ for various values of the exponent $\nu$, with parameters $b=0.95$, and $a=0.5$. In panel (b), the value $X_\mathrm{min}$ of the point of minimum of $\langle \tau_f\rangle$ is plotted as a function of the exponent $\nu$: the dashed vertical lines identify the values of $\nu$ used to plot, with the same color, the curves of $\langle \tau_f\rangle$ in (a), while the grey area  marks the range of $\nu\in(\nu_{1},\nu_{2})$ for which $X_{\min} = +\infty$, with $\nu_1(a=0.5, b=0.95)\simeq 0.89$, and $\nu_2(a=0.5,b=0.95)\simeq 1.32$.}\label{Fig:reset}
\end{figure}

\section{Periodic resetting and experimental protocol}\label{sec:RT_periodic}

In the previous sections, we have studied the first-passage properties of an RTP in the presence of Poissonian resetting 
where the time interval $T$ between two successive resettings is a random variable drawn from an exponential 
distribution $P(T) = r\,{\rm e}^{- r\,T}$. While this case is easier to study analytically, experimentally it is easier to 
implement a periodic protocol~\cite{Ciliberto,Faisant_2021} where the interval $T$ is fixed, and not a random variable. As in the
Brownian case, the details of the first-passage properties of the RTP in these two protocols differ from each other. However, the
qualitative behaviours of the MFPT as a function of the system parameters are similar, as in the Brownian case~\cite{Ciliberto, Faisant_2021}. Hence, we do not
present the details of these calculations with a periodic protocol here, but instead we outline below the salient features of the protocol and the main results.

The periodic protocol proceeds as follows. Initially, we let the dynamics of the RTP relax in a harmonic trap, with potential $V(x)=\kappa x^2/2$, for a time $T_{\rm eq}$. At the end of this \textit{equilibration} phase, the particle position is distributed according to Eq. \eqref{eq:rho}: this is true only if the typical relaxation time $\tau_{\rm rel}$ of the particle in the harmonic trap is much smaller than $T_{\rm eq}$, i.e., $\tau_{\rm rel} \ll T_{\rm eq}$. We recall that $T_{\rm eq}$ is the time interval during which the harmonic trap is switched on. The relaxation time $\tau_{\rm rel}$ of the RTP in this harmonic trap has been computed recently and it is given by $\tau_{\rm rel} = \kappa^{-1}$ \cite{Confpot}. At the end of this \textit{equilibration} phase, the confining potential is switched off and the particle resumes its free RTP dynamics for a given period $T$. During this \textit{search} phase, we keep track of the FPT statistics to a target at position $L$. After a time $T$, a new equilibration phase starts and measurements on the system are suspended. The process goes on by alternating search and equilibration phases (see Fig. \ref{Fig:expBM}). Thus, effectively, the motion of the RTP consists of a periodic resetting after a time $T$ to points drawn from the probability density $p(x)$ in Eq. (\ref{eq:rho}). As for the case of Brownian particles, it can be shown that the mean first-passage time $\langle t_f\rangle$ is given by  \cite{Ciliberto,Faisant_2021}
\begin{equation}\label{eq:tfT}
\langle t_f\rangle=\frac{\int_0^T \mathrm{d}\tau \,S(\tau)}{1-S(T)},
\end{equation}
where $S(t)=\int_{-\infty}^{+\infty}\mathrm{d}x\,p(x)S_0(t|x)$ is the survival probability of the same process without resetting, and initial position distribution $p(x)$ in Eq. (\ref{eq:rho}). As mentioned before, the results for the MFPT obtained by analysing Eq. (\ref{eq:tfT}) turn out to be qualitatively similar to the Poissonian resetting case, for which Eq. (\ref{eq:tr}) was the relevant formula. Effectively, the rate $r$ in the Poissonian resetting plays the same role as $1/T$ in the periodic resetting.



As in the case of Poissonian resetting, one of the main consequences of the periodic resetting is to make the MFPT of the RTP {\it finite}. We recall that in the absence of resetting, the MFPT of the RTP is {\it infinite}, since the FPT distribution has a power law decay $\propto t_f^{-3/2}$ for large $t_f$. This finite value is given by the formula (\ref{eq:tfT}) for periodic resetting. This formula is of course valid only when the system has relaxed to its true stationary state during the equilibration phase, i.e., when $T_{\rm eq} \gg \tau_{\rm rel} = \kappa^{-1}$. Hence, for experimental measurements of the MFPT, it is first important to verify whether this condition is satisfied of not. Note that it $T_{\rm eq} \ll \tau_{\rm rel}$, which means that the RTP rarely resets, the MFPT should diverge. Hence, the MFPT $\langle t_f \rangle$, when plotted as a function of the ratio $T_{\rm eq}/\tau_{\rm rel}$, is expected to diverge when the ratio goes to $0$ and to approach a constant value given by Eq. (\ref{eq:tfT}) when this ratio approaches $+\infty$. The measured data in experiments will correspond to the theoretical results discussed here only when the curve has flattened. 
We have performed numerical simulations with this periodic protocol and measured $\langle t_f \rangle$ as a function of this ratio $T_{\rm eq}/\tau_{\rm rel} = \kappa T_{\rm eq}$, as shown by the symbols in Fig. \ref{Fig:exp}. As expected, we see that this curve converges to a constant value, as the ratio $T_{\rm eq}/\tau_{\rm rel}  \to \infty$. Finally, we also computed the asymptotic value 
from the theoretical result in Eq. (\ref{eq:tfT}), as shown by the dashed horizontal lines. As seen in Fig. \ref{Fig:exp}, the agreement between simulations and the theory is excellent.


\begin{figure}[t]
\centering
\includegraphics[width = 0.8 \linewidth]{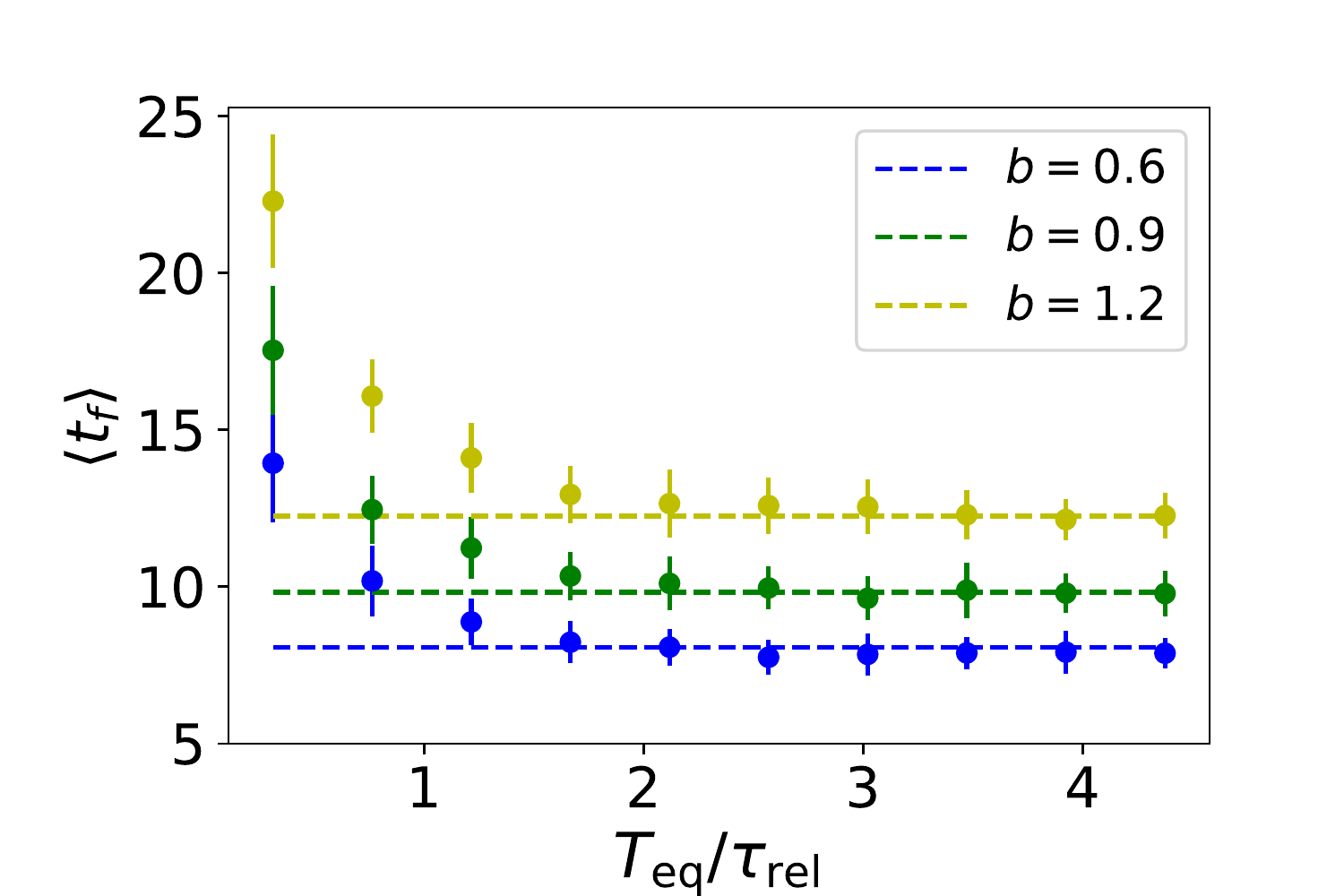}
\caption{Numerical estimates (symbols) of the the mean first-passage time as a function of the equilibration time $T_{\rm eq}/\tau_{\rm rel}$, for different values of $b=L/c$. The dashed horizontal line corresponds to the prediction of Eq. \eqref{eq:tfT}, i.e., with full realization. The values of $\langle t_f\rangle$ have been computed numerically via $N=2.5\times 10^4$ realizations of the process $x(t)$ with time step $\Delta t= 0.005$, with parameters $\tau_{\rm rel }=0.5$, $v=5$, $\gamma=2.5$, $c=v/\kappa=2.5$, $T=35$ and $\nu=\gamma/\kappa=1.25$. Error bars are given by standard deviation.}\label{Fig:exp}
\end{figure}



\section{Conclusions}\label{sec:Concl}

In this work, we have studied how the statistics of the first-passage time for RTP's, in one spatial dimension, to a target at position $L$ is modified by the introduction of two extra ingredients to their dynamics. First we 
studied the first-passage time distribution of a free RTP, without any resetting or any confining potential, but starting from an initial position drawn from an arbitrary distribution $p(x)$. For experimental purposes, it is relevant to consider the initial distribution $p(x)$ to be the stationary distribution of the RTP in the presence of a harmonic trap. This stationary distribution turns out to be non-Boltzmann and its shape is tunable by a parameter $\nu >0$. This parameter $\nu$ depends on the microscopic parameters of the RTP dynamics. 
In this particular case, we have shown that the first-passage time distribution of the free RTP, averaged over this initial distribution, develops interesting singularities depending on
the parameter $\nu$. This is summarised in Fig. \ref{Fig3}. 

In the second part of this work, we studied the mean first-passage time (MFPT) of an RTP subjected to stochastic resetting. In particular, the experimental protocol suggests to
study the MFPT for the case where the resetting position is distributed according to the non-Boltzmann stationary distribution $p(x)$ mentioned above, parametrised by $\nu>0$. Then, 
we first focused on the diffusive limit of the RTP and, by varying the parameter $\nu$ and the scaled target location $b = L/c$, we found a very rich phase diagram in $(b,\nu)$ plane, with a reentrance phase transition. In one phase (which we refer to as metastable), the scaled MFPT $\langle \tau_f \rangle$ as a function of the scaled resetting rate $X$, exhibits a non-monotonic decay, with a local minimum or a kink at $X_{\min}$. In the second phase (referred to as monotonic), the corresponding decay is monotonic. In the $(b, \nu)$ plane, there are two phase boundaries $\nu_1(b)$ and $\nu_2(b) > \nu_1(b)$ such that, for a fixed $b$, the phase is metastable for $\nu<\nu_1(b)$, monotonic for $\nu_1(b)<\nu<\nu_2(b)$ and metastable {\it again} for $\nu > \nu_2(b)$, as shown schematically in Fig. \ref{Fig_phdiag}. In the case of a more general RTP dynamics, i.e., beyond the diffusive limit, we found a qualitatively similar, yet somewhat richer behavior of $\langle \tau_f\rangle$ vs $X$ (see Fig. \ref{Fig:reset}). Finally, we have discussed the case of periodic resetting (see Fig. \ref{Fig:exp}).  

The results presented here are expected to be useful for possible future experiments for non interacting RTP's in the presence of a stochastic resetting, implemented by the thermal relaxation
in a harmonic trap.

\newpage

\appendix

\vspace*{1cm}

\section{Derivation of Eq. (\ref{Eq_Sr})} \label{sec:appren}

We consider an RTP trajectory, starting at the initial position $x_0$, and resetting to the new positions $\{x_1, x_2, \cdots \}$ after successive resettings,
where $\{x_0, x_1, x_2, \cdots \}$ are independent and identically distributed random variables, each drawn from $p(x)$. Let $S_r(t,\{x_i\})$ denote the joint probability 
that the particle survives up to time $t$ and the resetting positions take the values $\{x_0, x_1, x_2, \cdots \}$. We note that in a fixed time $t$, there can be no resetting event,
or one resetting event, two resetting events, etc. Hence, one can write a renewal equation using the fact that the intervals between successive resetting events are statistically independent
\begin{widetext}
\begin{eqnarray} \label{app_ren}
S_r(t,\{x_i\}) &=& {\rm e}^{-r t} S_0(t|x_0) + r \int_0^\infty {\rm d} \tau_1 {\rm e}^{-r \tau_1} \, S_0(\tau_1|x_0)\, S_0(t-\tau_1|x_1) \\
&+& r^2 \int_0^\infty {\rm d} \tau_1  \int_0^\infty {\rm d} \tau_2 \, {\rm e}^{-r \tau_1 - r \tau_2} S_0(\tau_1|x_0) S_0(\tau_2|x_1)  S_0(t-\tau_1-\tau_2|x_2) + \cdots \;,
\end{eqnarray}
\end{widetext} 
where $S_0(t|x_i)$ denotes the survival probability of the RTP up to time $t$, starting at $x_i$, and without any resetting. This Eq. (\ref{app_ren}) can be understood very simply. The first term corresponds to the event when there is no resetting up to time $t$. The second term corresponds to the event where there is only one resetting at time $0\leq \tau_1\leq t$. Here $x_1$ is the new starting point after the resetting event. Similarly the third term corresponds to the event with exactly two resettings in time $t$, respectively at time $\tau_1$ and $\tau_2$ and with $x_1$ and $x_2$ denoting the positions after the two resettings respectively. The dots in (\ref{app_ren}) represent the events with three, four, etc number of resettings in time $t$ and all these terms have a similar convolution structure. 

We first take the average of this relation (\ref{app_ren}) over the $x_i$'s, each drawn independently from $p(x)$. Let $S_r(t)$ denote the averaged survival probability with resetting and $S_0(t)$ the averaged survival probability without resetting. This gives, from Eq. (\ref{app_ren}),
\begin{widetext}
\begin{eqnarray} \label{app_ren_av}
S_r(t) &=& {\rm e}^{-r t} S_0(t) + r \int_0^\infty {\rm d} \tau_1 {\rm e}^{-r \tau_1} \, S_0(\tau_1)\, S_0(t-\tau_1) \\
&+& r^2 \int_0^\infty {\rm d} \tau_1  \int_0^\infty {\rm d} \tau_2 \, {\rm e}^{-r \tau_1 - r \tau_2} S_0(\tau_1) S_0(\tau_2)  S_0(t-\tau_1-\tau_2) + \cdots \;,
\end{eqnarray}
\end{widetext} 
where
\begin{eqnarray}\label{av_S0}
S_0(t) = \int {\rm d}x_0\, S_0(t|x_0)\, p(x_0) \,  \;.
\end{eqnarray}
The convolution structure in Eq. (\ref{app_ren_av}) naturally leads us to take the Laplace transform with respect to time $t$. We define
\begin{eqnarray}
\tilde S_r(s) &=& \int_0^\infty {\rm d}t\, S_r(t) {\rm e}^{-st}  \;, \label{LapSr_app} \\
\; \tilde S_0(s) &=& \int_0^\infty {\rm d}t\, S_0(t) {\rm e}^{-st} \label{LapS0_app} \;.
\end{eqnarray}
Taking the Laplace transform of Eq. (\ref{app_ren_av}) then gives the desired result
\begin{eqnarray}
\tilde{S}_r(s)=\frac{\tilde{S}_0(s+r)}{1-r\tilde{S}_0(s+r)} \;,
\end{eqnarray}
given in Eq. (\ref{Eq_Sr}) in the text. 

\section{Derivation of Eq.~(\ref{rho_bessel}).}\label{app_bessel}

In this Appendix, we provide the details of the derivation of the formula given in Eq. (\ref{rho_bessel}). Our starting point is the expression for $\rho_r(X,b)$ in Eq. (\ref{eq_rhor}) with $\rho(z)$ given
in Eq. (\ref{eq:rho}) which, for $b=1$, reads
\begin{eqnarray}\label{rho_bessel1}
\rho_r(X,1) = N(\nu) \int_{-1}^1 {\rm d}z\, (1-z^2)^{\nu-1} {\rm e}^{- X(1-z)}\,  \;,
\end{eqnarray} 
where we have used $|z-1| = (1-z)$ for $-1 \leq z \leq 1$ and where $N(\nu) = \Gamma(\nu+1/2)/(\Gamma(\nu)\sqrt{\pi})$. To make progress, we expand the exponential in Eq. (\ref{rho_bessel1}) in power series to get
\begin{equation}\label{rho_bessel2}
\rho_r(X,1) = N(\nu) \sum_{n=0}^\infty \frac{(-X)^n}{n!}  \int_{-1}^1 {\rm d}z\, (1-z^2)^{\nu-1} (1-z)^n  \;.
\end{equation}
The integral over $z$ can then be performed term by term, i.e.,
\begin{equation}\label{rho_bessel3}
 \int_{-1}^1 {\rm d}z \, (1-z^2)^{\nu-1} (1-z)^n  = 2^{2\nu+n-1} \frac{\Gamma(\nu+n)\Gamma(\nu)}{\Gamma(2\nu+n)} \;.
\end{equation}
Inserting this result in Eq. (\ref{rho_bessel2}) and rearranging, one obtains
\begin{equation} \label{rho_bessel4}
\rho_r(X,1) = \frac{2^{2\nu-1}}{\sqrt{\pi}} \Gamma\left(\nu+\frac{1}{2}\right) \sum_{n=0}^\infty (-2X)^n \frac{\Gamma(\nu+n)}{\Gamma(2\nu +n) n!}\;.
\end{equation}
Finally, the sum over $n$ can be performed explicitly leading to the result given in Eq. (\ref{rho_bessel}).

\nocite{*}
\bibliography{apssamp}

\end{document}